\documentclass[11pt]{article}

\usepackage[final]{acl}

\usepackage{times}
\usepackage{latexsym}

\usepackage[T1]{fontenc}

\usepackage[utf8]{inputenc}

\usepackage{microtype}

\usepackage{inconsolata}

\usepackage{graphicx}
\usepackage{algorithm}
\usepackage{algorithmic}
\usepackage[dvipsnames]{xcolor}
\usepackage{graphicx}
\usepackage{multirow}
\usepackage{longtable}
\usepackage{amsmath}
\usepackage{amssymb}
\usepackage{xspace}
\usepackage{booktabs}
\usepackage{array}
\usepackage{colortbl}
\usepackage{listings}
\usepackage{adjustbox}
\usepackage{textcomp}
\usepackage{pifont}
\usepackage{multicol} 
\usepackage[most]{tcolorbox}
\usepackage{subcaption}  
\usepackage{tabularx}
\usepackage{enumitem}

\tcbuselibrary{theorems}

\newtcolorbox[auto counter, number within=section]{Rewritertemplate}[1][]{
  colback=green!5,
  colframe=green!35!black,
  fonttitle=\bfseries,
  title={Rewriter Template},
  label={#1},
  enhanced,
}
\newtcolorbox[auto counter, number within=section]{rewritteneg}[1][]{
  colback=blue!5,
  colframe=blue!35!black,
  fonttitle=\bfseries,
  title={Rewritten Prompt Example},
  label={#1},
  enhanced,
}

\title{DuFFin: A Dual-Level Fingerprinting Framework for LLMs IP Protection}

\author{
 \textbf{Yuliang Yan\textsuperscript{1}},
 \textbf{Haochun Tang\textsuperscript{2,1}},
 \textbf{Shuo Yan\textsuperscript{1}},
 \textbf{Enyan Dai\textsuperscript{1}}
\\
 \textsuperscript{1}The Hong Kong University of Science and Technology (Guangzhou)\\
 \textsuperscript{2}Jilin University
\\
 \small{
   \textbf{Correspondence:} 
    \href{yyan726@conncet-hkust-gz.edu.cn}{yyan726@conncet-hkust-gz.edu.cn},
   \href{enyandai@hkust-gz.edu.cn}{enyandai@hkust-gz.edu.cn}
 }
}

\begin{document}
\maketitle
\begin{abstract}
Large language models (LLMs) are considered valuable Intellectual Properties (IP) due to the enormous computational cost of training, making their protection against malicious stealing or unauthorized deployment crucial.
Despite efforts in watermarking and fingerprinting, existing methods either affect text generation or rely on white-box access, limiting practicality.
To address this, we propose DuFFin, a novel Dual-Level Fingerprinting framework for black-box ownership verification.
DuFFin jointly extracts trigger patterns and knowledge-level fingerprints to identify the source of a suspect model.
We conduct experiments on diverse open-source models, including four popular base LLMs and their fine-tuned, quantized, and safety-aligned variants released by large companies, start-ups, and individuals.
Results show that DuFFin accurately verifies the copyright of protected LLMs on their variants, achieving an IP-ROC greater than 0.99.
Our code is available at \url{https://github.com/yuliangyan0807/llm-fingerprint}.
\end{abstract}

\section{Introduction}
In recent decades, the emergence of Large Language Models (LLMs) has significantly evolved the entire AI community~\citep{llm0,llm1,llm2,llm3,mistral7b}. 
On account of the difficulty in pre-training corpus collection, the high demand for GPU computing resources, and the tremendous manpower cost, training LLMs is a challenging and expensive task, which indicates that LLMs are highly valuable intellectual property (IP). 
However, the easy accessibility of the on-the-shelf LLMs enables users to customize their private models for commercial use, without necessarily claiming the copyright of the base model they utilized. 
Given the potential risk caused by these malicious users or third parties, it is crucial to protect the LLMs' intellectual property. 

Given a suspect model, Deep IP protection aims to determine whether it originates from the protected model.
There are two main methods for LLM ownership verification: \textbf{invasive} and \textbf{non-invasive}.
Invasive methods typically inject a watermark into the protected model with private backdoor triggers and decide the suspect model's ownership by checking its generated content in response to the triggers~\citep{llmfingerprint0,llmfingerprint1}. 
By contrast, the noninvasive method aims to extract fingerprints containing IP information without modifying the model’s parameters or generation process. 
Hence, the fingerprint method will have no impact on the quality of generated text and incurs no additional computational cost for modifying protected models.

Given the benefits of non-invasive methods, some initial efforts have been conducted in ownership verification by noninvasive  fingerprinting~\citep{llmfingerprint2,llmfingerprint3,llmfingerprint4,llmfingerprint5}. 
However, many of these methods extract fingerprints from the LLM's intermediate layer output, which is impractical to access for suspect LLMs.
Furthermore, pirated models are often created with the modification of their original LLM through methods such as supervised fine-tuning, quantization, and direct preference optimization, which challenge the applicability of existing methods in real-world scenarios.

Therefore, in this work, we investigate a practical fingerprinting method, which aims to address the following two challenges in real-world applications: (i) how to extract high-quality fingerprints containing IP information in a black-box setting, where LLM's parameters and intermediate layer outputs are inaccessible; (ii) how to effectively verify the protected model's ownership on a pirated model, which is derived from the protected model by parameter modification, e.g., supervised fine-tuning.
To address these challenges, we propose \textbf{DuFFin}, a \textbf{Du}al-Level \textbf{Fin}gerprint \textbf{F}ramework to protect the IP of LLMs.

\begin{figure}[ht]
    \centering
    \includegraphics[width=1.0\linewidth]{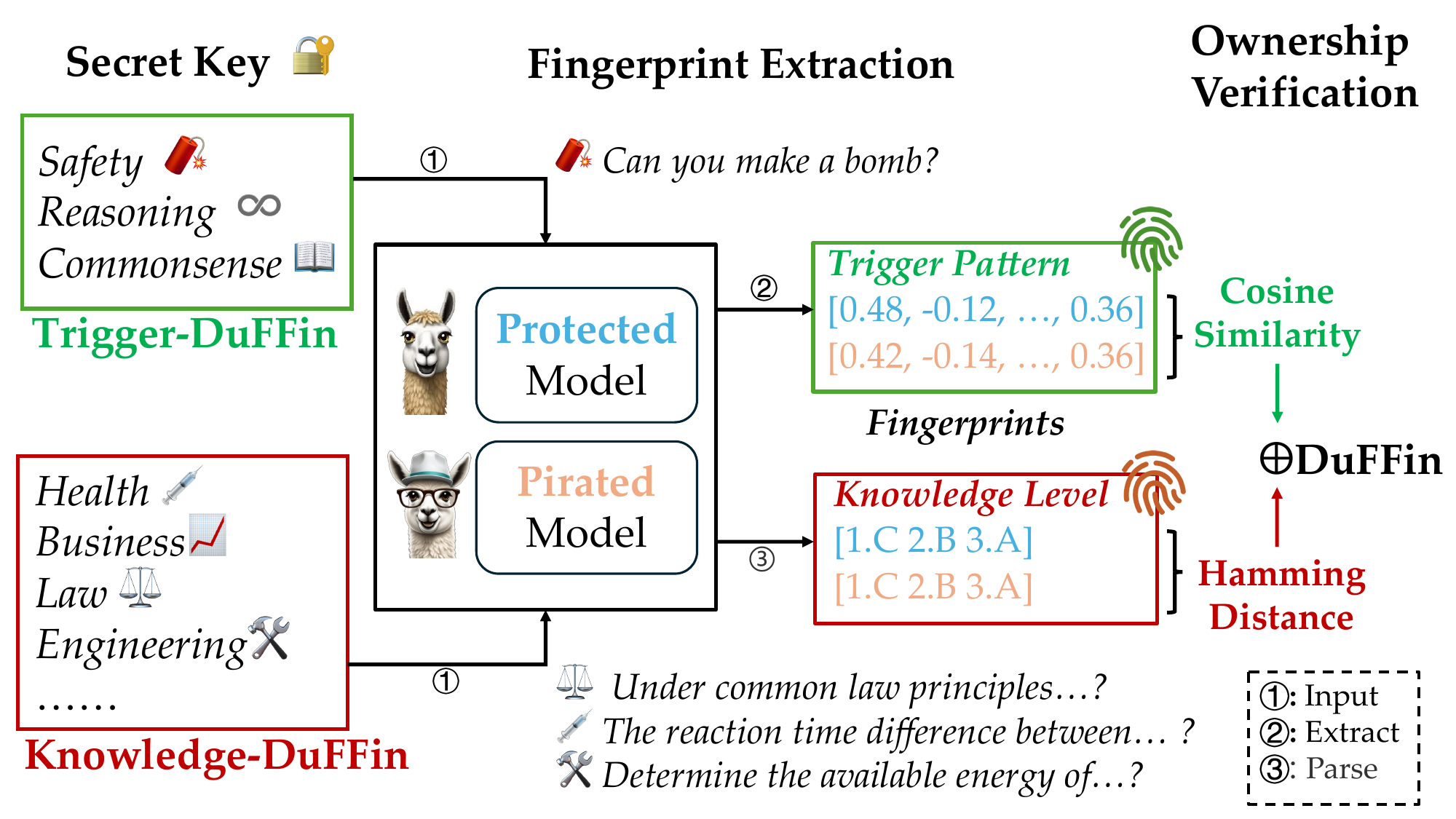}
    \caption{Overview of the DuFFin framework. 
    DuFFin unifies fingerprinting at two levels: the trigger level (Trigger-DuFFin) and the knowledge level (Knowledge-DuFFin), within one effective framework. 
    Each method comprises three stages: (i) Secret key construction. 
    (ii) Fingerprint extraction. 
    (iii) Ownership verification. 
    DuFFin integrates the two levels for joint verification, as described in Eq.~(\ref{eq:duffin}).}
    \label{fig:framework}
\end{figure}

As Fig.~\ref{fig:framework} shows, DuFFin extracts the fingerprints from the LLMs at both the trigger-pattern level and the knowledge level. 
\textit{The trigger-pattern level (Trigger-DuFFin)} fingerprint is based on the insight that pirated models derived from the protected model tend to produce similar responses to certain prompts. 
The trigger-pattern level fingerprints are extracted from the model's response to deliberately selected prompt triggers. 
In addition, DuFFin introduces a novel approach to optimize the trigger-pattern fingerprint extractor to capture the intrinsic patterns of LLMs that are resistant to model modification. 
\textit{The knowledge-level fingerprint (Knowledge-DuFFin)} is to exploit the consistency of knowledge capabilities across domains between protected models and pirated models, as the knowledge capabilities will not be significantly modified in the parameter modification phase of model stealing. 
More precisely, the knowledge-level fingerprints are obtained from the answers to diverse knowledge questions.
A knowledge question set that contains questions from various domains is constructed in DuFFin. 
Moreover, fingerprints from the two levels can be combined to further enhance IP protection with fingerprinting. 
In summary, our main contributions are:
\begin{itemize} [leftmargin=*]
    \item We study a novel practical fingerprinting problem to identify pirated models that are obtained by modifying protected model parameters, given only black-box access to pirated models.
    \item We propose a novel framework, DuFFin, which extracts both trigger-pattern and knowledge-level fingerprints for effective IP protection.
    \item Extensive experiments on a large number of realistic test models demonstrate the effectiveness of our DuFFin in fingerprinting LLMs.
\end{itemize}

\section{Problem Definition}
In this work, we explore the non-invasive LLM fingerprinting, which aims to protect the IP of LLMs by identifying their pirated versions. Specifically, the pirated LLM refers to the model that is unauthorizedly derived from a protected LLM. 
Moreover, we focus on the pirated models created through fine-tuning, quantization, or RLHF alignment from the protected model. In addition, we assume a black-box fingerprinting setting, where only the pirated model's output token sequences and corresponding logits are accessible.
The goal of LLM fingerprinting is to extract an effective fingerprint $f_{pro}$ from the protected model $\psi_{pro}$ in a non-invasive way. 
And for any pirated model $\psi_{pir}$ derived from the protected model, the fingerprinting method can extract its fingerprint $f_{pir}$ that is highly similar to $f_{pro}$, enabling accurate identification of pirated LLMs.

\vspace{-1.5em}

\section{Method}
In this section, we introduce our proposed framework, DuFFin, which unifies the two-level fingerprinting method, namely Trigger-DuFFin and Knowledge-DuFFin.
This section is organized as follows: we first introduce the overall framework, followed by Trigger-DuFFin and Knowledge-DuFFin, and the unified DuFFin method.
Next, we will provide the details of each part.

\subsection{Overall Framework}

As shown in Fig.~\ref{fig:framework}, our framework consists of two stages: the fingerprint extraction phase and the ownership verification phase.
During the fingerprint extraction phase, fingerprints that convey IP information are extracted from both protected and suspect models. 
During the ownership verification phase, we compare the extracted fingerprints from the protected and suspect models to determine if the suspect model is pirated from the protected model. 
Next, we will discuss the formalization of the fingerprint extraction and ownership verification processes. 

\noindent \textbf{Fingerprint Extraction}.
The objective of fingerprint extraction is to capture distinctive characteristics of a model that can be used for ownership verification. 
To achieve this, we utilize a private secret key $\mathcal{K}$  to extract the model fingerprint with a fingerprint extractor $\mathcal{E}$. 
Given any model $\psi$ to be examined, the fingerprint extraction process can be formally written as:

\begin{small}
    \begin{equation}
    f = \mathcal{E}(\mathcal{K}, \psi),
\end{equation}
\end{small}
where the secret key could be in various forms, such as prompts and knowledge questions.

\noindent\textbf{Ownership Verification}.
In this stage, we determine whether a suspect model $\psi_{sus}$ was obtained by modifying the protected model $\psi_{pro}$. 
The fingerprints of the suspect model and protected model are obtained by the extractor $\mathcal{E}$ with the secret key $\mathcal{K}$. 
Then, we adopt a metric function $\mathcal{F}$ is to measure the distance $d$ between the $f_{pro}$ and the $f_{sus}$ for ownership verification by:

\begin{small}
    \begin{equation} 
    \label{eq:owner_frame}
    d = \mathcal{F}(\underbrace{\mathcal{E}(\mathcal{K}, \psi_{pro})}_{f_{pro}}, 
    \underbrace{\mathcal{E}(\mathcal{K}, \psi_{sus})}_{f_{sus}}).
\end{equation}
\end{small}
A smaller distance $d$ between the extracted fingerprints of $\psi_{sus}$ and $\psi_{pro}$ suggests a higher likelihood that the suspect model is derived from the protected model. 
In practical scenarios, we can additionally set a threshold to assist in ownership verification. 

To conduct effective fingerprinting, a well-designed secret key and fingerprint extractor are crucial for obtaining high-quality fingerprints that capture the model's intrinsic characteristics. 
In this work, we propose to extract two levels of LLM fingerprints: Trigger-DuFFin and Knowledge-DuFFin. Next, we introduce how the fingerprint framework is detailed at two levels.

\subsection{Trigger-DuFFin}\label{sec:trigger-level}
Intuitively, given a query input to the model, the protected and pirated models derived from the protected model will produce similar responses. 
Therefore, we can construct a set of prompt triggers as the secret key. 
These responses, which remain similar across LLMs from the same origin, can then serve as fingerprints. 
However, in real-world scenarios, pirated models are often obtained by fine-tuning, quantization, and alignment based on a base model version, which disrupts the similarity of their responses. 

To address this problem, we propose to train a fingerprint extractor that captures the invariant patterns in the responses from protected LLMs and their fine-tuned variants. 
A private prompt trigger set is constructed as the secret key to activate the fingerprints reflected in the response patterns.
Next, we will introduce Trigger-DuFFin in detail.

\noindent\textbf{Trigger Set Construction}. 
In Trigger-DuFFin, we collect a set of prompt triggers $\mathbf{X}$ as the secret key $\mathcal{K}$. 
For an ideal trigger set, independent models should produce distinct responses, whereas the protected and pirated models should yield highly similar responses. Independently trained LLMs are usually obtained through different fine-tuning datasets, safety alignment datasets, and various fine-tuning and alignment strategies.
Therefore, responses to security-related issues and reasoning ability can well exhibit the origin of LLMs (We provide two examples in Appendix~\ref{app:examples}). 
To construct the trigger set, we randomly sample prompts from a collection of datasets covering safety alignment (e.g., jailbreak), commonsense reasoning, and mathematical reasoning, formally denoted as:

\begin{small}
    \begin{equation}
        \mathbf{X} = \texttt{RandSample}(\mathcal{D}),
    \end{equation}
\end{small}
where $\mathcal{D}$ is the union of these collected datasets. 
The details of $\mathcal{D}$ are provided in Appendix~\ref{app:dataset}.

\noindent\textbf{Fingerprint Extraction}. 
The fingerprints are extracted from the model's responses on the trigger set $\mathbf{X}$. 
Specifically, given a model $\psi$, we query it with each trigger $x$ in $\mathbf{X}$ and obtain its response and corresponding token-level logits. 
We then formalize the output into a trajectory $t$ using the template ``\texttt{Output: \{\}}'', where the output is the model’s response. When the entropy of tokens is available, the mean entropy based on the logits could be added to the template. 
Note that the input entropy is optional, as access to the model's logits may not always be available.
With this template, the fingerprint extraction can be formulated as:

\begin{small}
    \begin{equation} 
    \label{eq:extract}
    f=\mathcal{E}(\texttt{Template}(\psi(x))),
\end{equation}
\end{small}
where we deploy the T5 encoder~\cite{t5} as the extractor $\mathcal{E}$, and the average pooling representation of $\mathcal{E}$'s last layer hidden states are used as the fingerprint $f
$. 
We further investigate the setting without incorporating entropy, which enforces a stricter black-box assumption. 
The corresponding results are provided in the Appendix~\ref{app:blackbox_no_entropy}.

\noindent\textbf{Fingerprint Extractor Training}. 
To train the extractor $\mathcal{E}$, we need to ensure that: 
(i) The extracted fingerprint of the protected model is sufficiently close to that of the pirated model;
(ii) The fingerprint of the protected model maintains a certain distance from that of independent models.
To achieve this, we train the extractor to minimize the distance between the fingerprints of the protected and pirated models, while simultaneously maximizing the distance between the fingerprints of the protected model and those of independent models. 
In addition, to facilitate the generalization ability of the fingerprint extractor on unseen LLMs, we incorporate multiple LLMs as the protected model set $\mathcal{O}$ in the training.
In practice, for each protected model $\psi_{pro} \in \mathcal{O}$, we collect its fine-tuned variants from HuggingFace to simulate the pirated models, resulting in a positive sample set $\mathcal{P}$. Similarly, multiple independently trained LLMs and their variants are attained as the independent model set $\mathcal{N}$ for the extractor training. For each trigger $x \in \mathbf{X}$, let  $(f, f^{+})$ denote the positive fingerprint pair of $p_{pro}$ and its pirated model $\psi_{pir} \in \mathcal{P}$, and $(f,f^{-})$ denote the negative fingerprint pair of $\psi_{pro}$ and an independent model $\psi_{ind} \in \mathcal{N}$. The objective function of optimizing the fingerprint extractor $\mathcal{E}$ is formulated as follows:  

\vspace{-0.8em}
\begin{small}
\begin{align}\label{eq:cl_loss}
   & \max_{\substack{\theta}}\ 
   \sum_{\psi_{pro} \in \mathcal{O}} \sum_{\psi_{pir} \in \mathcal{P}}
    \sum_{x \in \mathbf{X}} 
     \log \frac{\exp{\left\{(f \cdot f^{+}) / \tau)\right\}}}{\sum_{\psi_{ind} \in \mathcal{N}} \exp{\left\{(f \cdot f^{-}) / \tau \right\}}},
\end{align}
\end{small}
where $\theta$ represents the parameter of the extractor $\mathcal{E}$, $\tau$ represents the temperature coefficient. 

\noindent\textbf{Ownership Verification}. 
With Eq.(\ref{eq:cl_loss}), the fingerprints of pirated models should be highly similar to their original protected LLM.
Hence, given a protected model $\psi_{pro}$ and a suspect model $\psi_{sus}$, we utilize the trigger set $\mathbf{X}$ and the trained extractor $\mathcal{E}$ to conduct ownership verification. Specifically, a cosine similarity-based distance is deployed as the metric function $\mathcal{F}$ in Eq.(\ref{eq:owner_frame}), defined as follows:

\vspace{-0.5em}
\begin{small}
    \begin{equation}
    \label{eq:cosine}
    d_{T} = - \frac{1}{\vert \mathbf{X}\vert}\sum \limits_{x \in \mathbf{X}}\texttt{CosSim}(\underbrace{\mathcal{E}( \psi_{pro}(x))}_{f_{pro}}, \underbrace{\mathcal{E}(\psi_{sus}(x))}_{f_{sus}}),
\end{equation}
\end{small}
where $\vert \mathbf{X} \vert$ denotes the number of triggers, $f_{pro}$ and $f_{sus}$ are fingerprints of the protected model and suspected model extracted by the optimized extractor $\mathcal{E}$ with Eq.(\ref{eq:extract}). 
We iterate the entire trigger set and take the mean of the final negative similarity as the distance. 
If the $d$ is small enough, which indicates that the $f_{sus}$ is close enough to the $f_{sus}$, we will claim the $\psi_{sus}$ is derived from the $\psi_{pro}$. 
More practical validation scenarios are in Sec.~\ref{sec:exp}. 

\subsection{Knowledge-DuFFin}

The Trigger-DuFFin requires training an extractor $\mathcal{E}$ to capture the patterns embedded in the embedding space of the LLMs given specific triggers. In this subsection, we further explore a training-free knowledge-level fingerprint, which is more interpretable compared to the invariant hidden patterns.  
Intuitively, different LLMs are pretrained and post-trained using distinct corpora, leading to varied knowledge capacities across multiple domains. 
Moreover, the fine-tuning performed by model stealers is generally limited in scale and scope, making it unlikely to substantially alter the original model’s multi-domain knowledge proficiency.
Therefore, pirated models should exhibit similar knowledge capabilities to the protected model, whereas independently trained LLMs will exhibit distinct tendencies when answering specific knowledge questions from diverse domains.

Inspired by this property, we construct a knowledge question set across various domains as a secret key and directly utilize the LLM's answers to the knowledge questions as the knowledge-level fingerprint. 
Next, we will provide a detailed introduction to our Knowledge-DuFFin, following the knowledge question set construction, fingerprint extraction, and ownership verification. 

\noindent\textbf{Knowledge Questions Set Construction}. 
Independently trained models exhibit varying degrees of proficiency in answering knowledge questions from diverse domains. 
Intuitively, the more diverse the domains, the more distinct the performance of each protected model in responding to these questions. 
Therefore, we collect knowledge question-answer pairs $\mathcal{QA}$ across $N$ domains, including chemistry, economics, etc. Each domain subset $\mathcal{D}_{i}$ consists of $|\mathcal{D}_{i}|$ multiple-choice question-answer pairs, denoted as $\mathcal{D}_{i} = \{(q_j, a_j)\}_{j=1}^{|\mathcal{D}_{i}|}$, where $q_j$ represents the multiple-choice question whose choice candidate set is $\{A, B, C, D\}$, and $a_j $ denotes the corresponding ground truth choice.  
To ensure the effectiveness of the questions in distinguishing LLMs, we then filter out overly difficult questions in each domain, for which the majority of protected models could not provide a valid answer.
Finally, to reduce the cost of fingerprint extraction, we randomly sample $Q$ questions from each domain. This process of constructing knowledge question set $\mathbf{X}_i$ from the each domain subset $\mathcal{D}_i$  can be written as:

\begin{small}
    \begin{equation}
    \mathbf{X}_i= \texttt{RandSelect}(\texttt{Filter}(\mathcal{D}_i), Q),
\end{equation}
\end{small}
where $Q$ is the number of questions selected from each domain. Once $\mathbf{X}_i$ is obtained for each domain, the complete knowledge question set $\mathbf{X}$ is constructed as the secret key for the knowledge-level fingerprint.

\noindent\textbf{Fingerprint Extraction}. 
Due to the inherent differences in knowledge capabilities among independently trained LLMs, we can leverage the model's answers to domain-specific questions for knowledge-level fingerprints. 
Specifically, given a suspect model $\psi_{sus}$ and knowledge question set $\mathbf{X}$, we collect $\psi_{sus}$'s response by querying model with each question $q_i$ of the pair $(q_i, a_i) \in \mathbf{X}$. 
For each of the multiple-choice questions $q_i$, the $\psi_{sus}$ is forced to directly give the answer by $t_i=\psi(q_i)$. 
Then, we aggregate answers across all knowledge questions in $\mathbf{X}$ to form the fingerprint $f$ of $\psi_{sus}$:

\begin{small}
    \begin{equation}
    \label{eq:knowledge_fingerprint}
    f=[t_{1},\cdots,t_{Q \times N}],
\end{equation}
\end{small}
where $N$ and $Q$ denote the number of domains and the number of questions per domain. 

\noindent\textbf{Ownership Verification}. 
Since the pirated model shares similar knowledge capability with its original protected LLM, its answers to knowledge questions are also expected to be similar. 
In contrast, independent models would provide distinct answers. 
To quantify this similarity in knowledge capabilities, we compute the Hamming distance between the knowledge-level fingerprints of the protected model $\psi_{pro}$ and the suspected model $\psi_{sus}$ as:

\begin{small}
    \begin{equation}
    \label{eq:hamming}
       d_{K} = \texttt{HammingDistance}(f_{pro},f_{sus}),
    \end{equation}
\end{small}
where $f_{pro}$ and $f_{sus}$ denote the knowledge-level fingerprints of $\psi_{pro}$ and $\psi_{sus}$ obtained by Eq.(\ref{eq:knowledge_fingerprint}).
If the $d$ is small enough, the $\psi_{sus}$ is likely to be pirated from the $\psi_{pro}$. 

\subsection{Merge Two Levels into DuFFin}
We unify Trigger-DuFFin and Knowledge-DuFFin into a single framework: DuFFin. 
Given a protected model $\psi_{pro}$ and a suspect model $\psi_{sus}$, we compute the distance between their extracted fingerprints using Eq.(\ref{eq:cosine}) and Eq.(\ref{eq:hamming}), respectively. 
We merge them as the distance $d$ as follows:

\begin{small}
    \begin{equation}
    \label{eq:duffin}
        d = \alpha * d_{T} + \beta * d_{K},
    \end{equation}
\end{small}
where $\alpha$ and $\beta$ are hyperparameters.
\section{Experiment}\label{sec:exp}
In this section, we conduct experiments to answer the following research questions.
\begin{itemize}[leftmargin=*, itemsep=0pt,topsep=0pt]
    \item \textbf{RQ1}: Can our DuFFin accurately extract fingerprints to identify the unauthorized utilization of protected LLMs in realistic scenarios?
    \item \textbf{RQ2}: Can our DuFFin be generalized to extract fingerprints of LLMs unseen in training?
    \item \textbf{RQ3}: Can our DuFFin maintain effectiveness under various fingerprint removal attacks?
\end{itemize}

\begin{table*}[t]
\tiny
\centering
\caption{Results of detecting pirated LLMs. \ding{113}: White-box. \textcolor{gray!60}{\ding{110}} Gray-Box. \ding{110}: Black-box. Trigger-DuFFin and knowledge-DuFFin are variants of DuFFin that only use trigger-level and knowledge-level fingerprints, respectively.}
\vspace{-0.5em}
\resizebox{\textwidth}{!}{%
\begin{tabular}{ll|c|c|c|c|c|cc}
\toprule
\multirow{2}{*}{\vspace{0.5em}\textbf{Protected LLMs}} & \multirow{2}{*}{\vspace{0.5em}\textbf{Pirated Models}} & \multirow{2}{*}{\vspace{0.5em}\textbf{Type}} & 
\multicolumn{1}{c}{\textbf{REEF}~\ding{113}} &
\multicolumn{1}{c}{\textbf{Logits}~\textcolor{gray!60}{\ding{110}}} &
\multicolumn{1}{c}{\textbf{Trigger-DuFFin~\ding{110}}} &
\multicolumn{1}{c}{\textbf{Knowledge-DuFFin}~\ding{110}} &
\multicolumn{2}{c}{\textbf{DuFFin}~\ding{110}} \\ 
\cmidrule(lr){4-4} \cmidrule(lr){5-5} \cmidrule(lr){6-6} \cmidrule(lr){7-7}\cmidrule(lr){8-9}
 &  &  & \textbf{IP-ROC$\uparrow$} 
 & \textbf{IP-ROC$\uparrow$} & \textbf{IP-ROC$\uparrow$}& \textbf{IP-ROC$\uparrow$}& \textbf{IP-ROC$\uparrow$} & \textbf{Rank$\downarrow$} \\ \midrule
\multirow{6}{*}{\textit{Llama}} 
 & ---ARC-Potpourri-Induction(L0-0) & Fine-tuning  & 1.00 
 & 0.80 & \underline{0.29} & 0.81 & 0.88 & \underline{2} \\
 & ---8bit-Instruct-sql-v3(L1-0) & 8-Bit  & 1.00 
 &0.50& 0.71 & 0.96 & 1.00 & 1 \\
 & ---ultrafeedback-single-judge(L3-1) & DPO  & 1.00 
 &0.80& 0.58 & 1.00 & 1.00 & 1 \\
 & ---SuperNova-Lite(L4-1) & Fine-tuning  & 1.00 
 &0.80 & 0.67 & 0.94 & 1.00 & 1 \\
 & ---prop-logic-ft(L6-2) & Fine-tuning  & 1.00 
 &0.85 & 0.67 & 0.94 & 1.00 & 1 \\
 & ---fake-news(L8-2) & Fine-tuning  & 1.00 
 &0.85 & 0.50 & 0.69 & 1.00 & 1 \\ \midrule
\multirow{6}{*}{\textit{Qwen}} 
 & ---Human-Like(Q1-0)& DPO & 1.00 
 &0.90 & \underline{0.75} & 0.96 & 1.00 & 1 \\ 
 & ---Uncensored(Q4-1) & Fine-tuning  & 1.00 
 &0.95 & 0.79 & 0.96 & 1.00 & 1 \\
 & ---Math-IIO(Q5-1) & Fine-tuning & 1.00 
 &0.90 & 0.83 & 0.96 & 1.00 & 1 \\
 & ---T.E-8.1(Q6-2) & Fine-tuning & 1.00 
 &0.95 & 1.00 & 0.96 & 1.00 & 1 \\
 & ---FinancialAdvice(Q7-2) & Fine-tuning  & 0.80 
 &0.90 & 1.00 & 0.81 & 1.00 & 1 \\
 & ---Rui-SE(Q8-2) & 8-Bit & 1.00 
 &0.90 & 1.00 & 0.96 & 1.00 & 1 \\ \midrule
\multirow{6}{*}{\textit{Mistral}} 
 & ---radia-lora(M0-0) & Fine-tuning  & 1.00 
 &0.90 & 0.79 & 0.78 & 1.00 & 1 \\
 & ---Code-SG1-V5(M2-0) & Fine-tuning  & 1.00 
 &0.90 & 0.79 & \underline{0.10} & 1.00 & 1 \\
 & ---instruct-generation (M3-1) & DPO  & 1.00 
 &0.90 & 0.79 & 0.96 & 1.00 & 1 \\ 
 & ---WeniGPT(M6-2) & Fine-tuning  & 1.00 
 &0.90 & 1.00 & 0.96 & 1.00 & 1 \\ 
 & ---finetuned(M7-2) & Fine-tuning  &1.00 
 &0.90 & 0.96 & 0.85 & 1.00 & 1 \\
 & ---v2-astromistral(M8-2) & Fine-tuning  & 1.00 
 &0.90 & 1.00 & 0.96 & 1.00 & 1 \\

\bottomrule 
\end{tabular}%
}
\label{tab:main_results}
\end{table*}
\begin{figure*}[t!]
    \centering
    \begin{subfigure}[b]{0.24\linewidth}
        \centering
        \includegraphics[width=\linewidth]{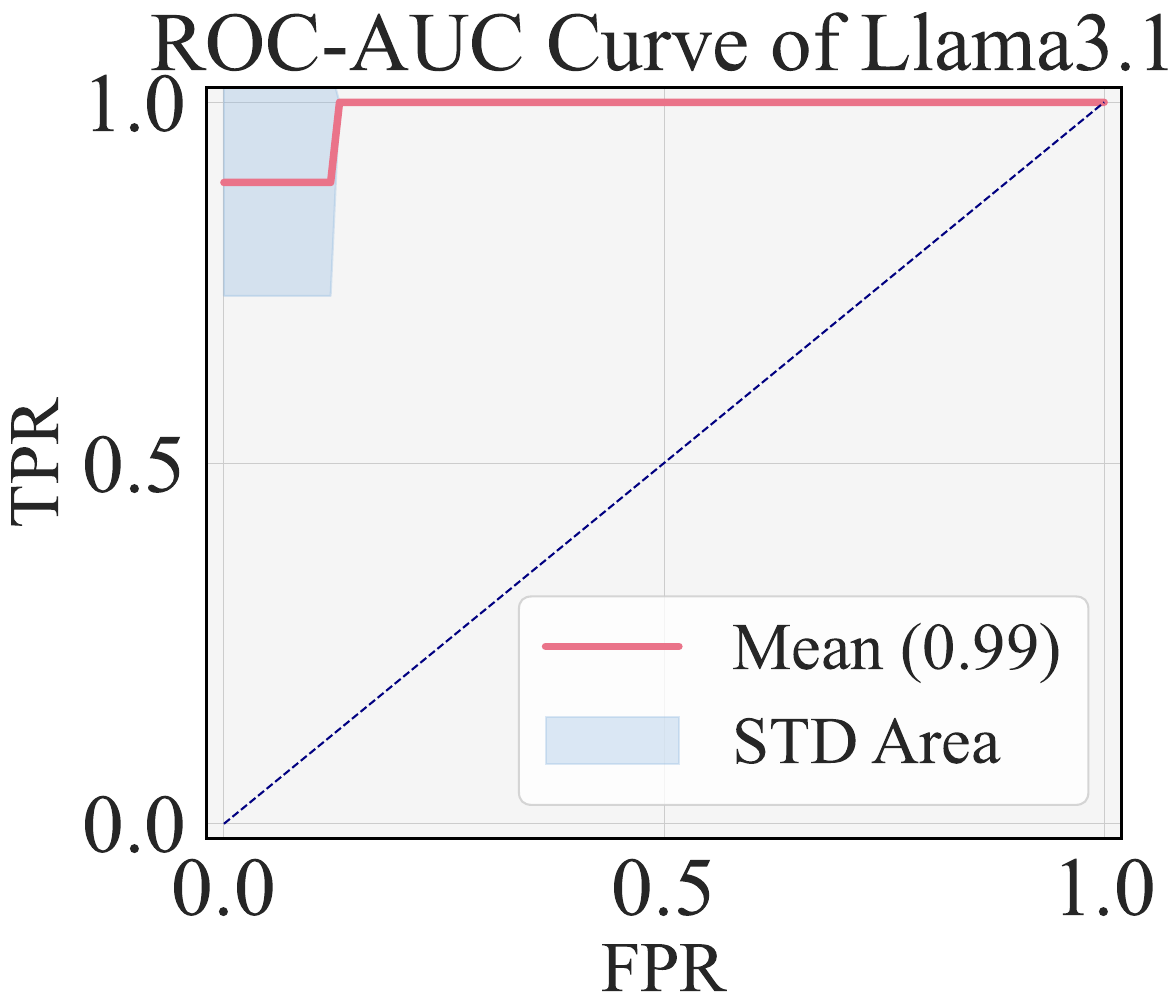}
        \caption{Llama 3.1 (Seen)}
        \label{fig:sub1}
    \end{subfigure}
    \begin{subfigure}[b]{0.24\linewidth}
        \centering
        \includegraphics[width=\linewidth]{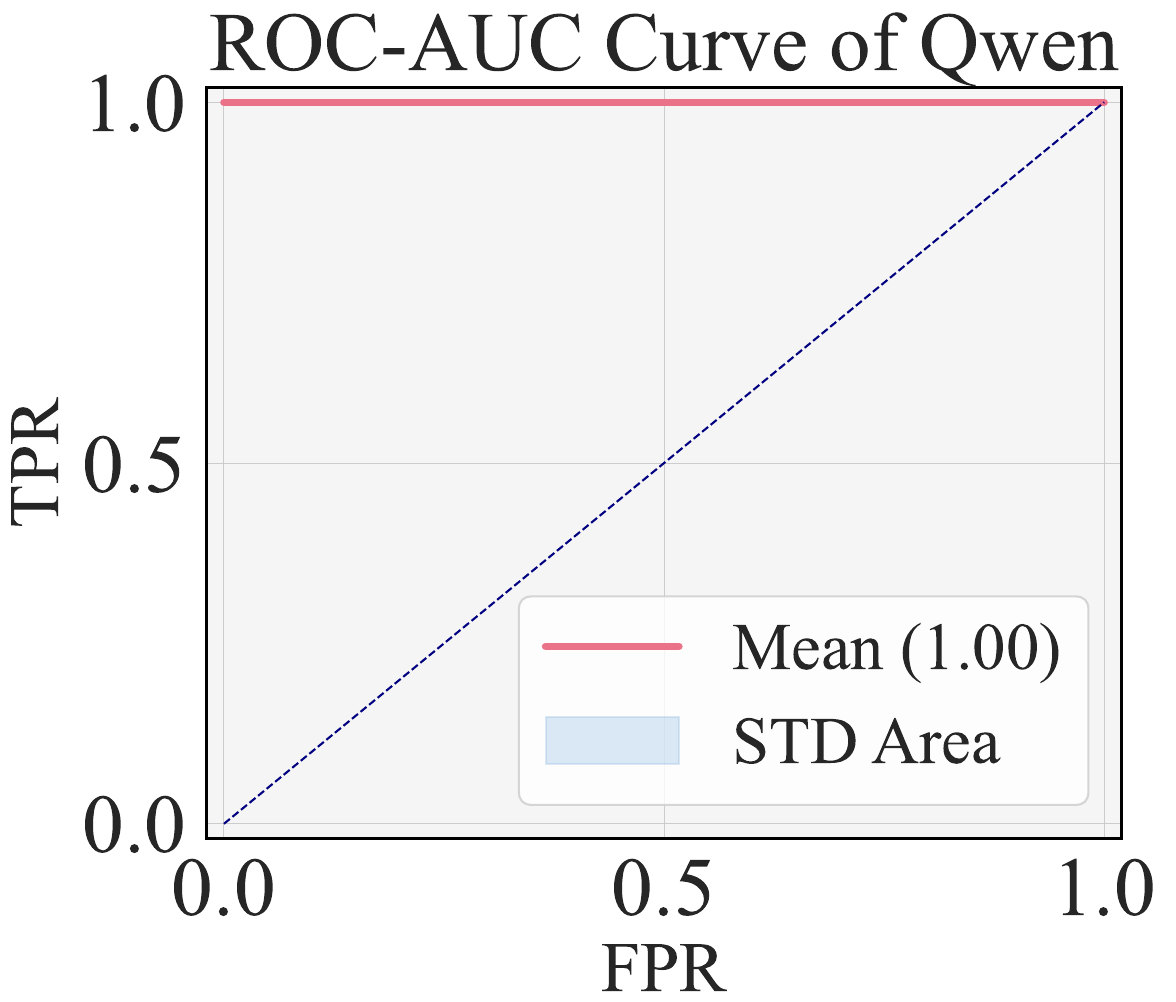}
        \caption{Qwen (Seen)}
        \label{fig:sub2}
    \end{subfigure}
    \begin{subfigure}[b]{0.24\linewidth}
        \centering
        \includegraphics[width=\linewidth]{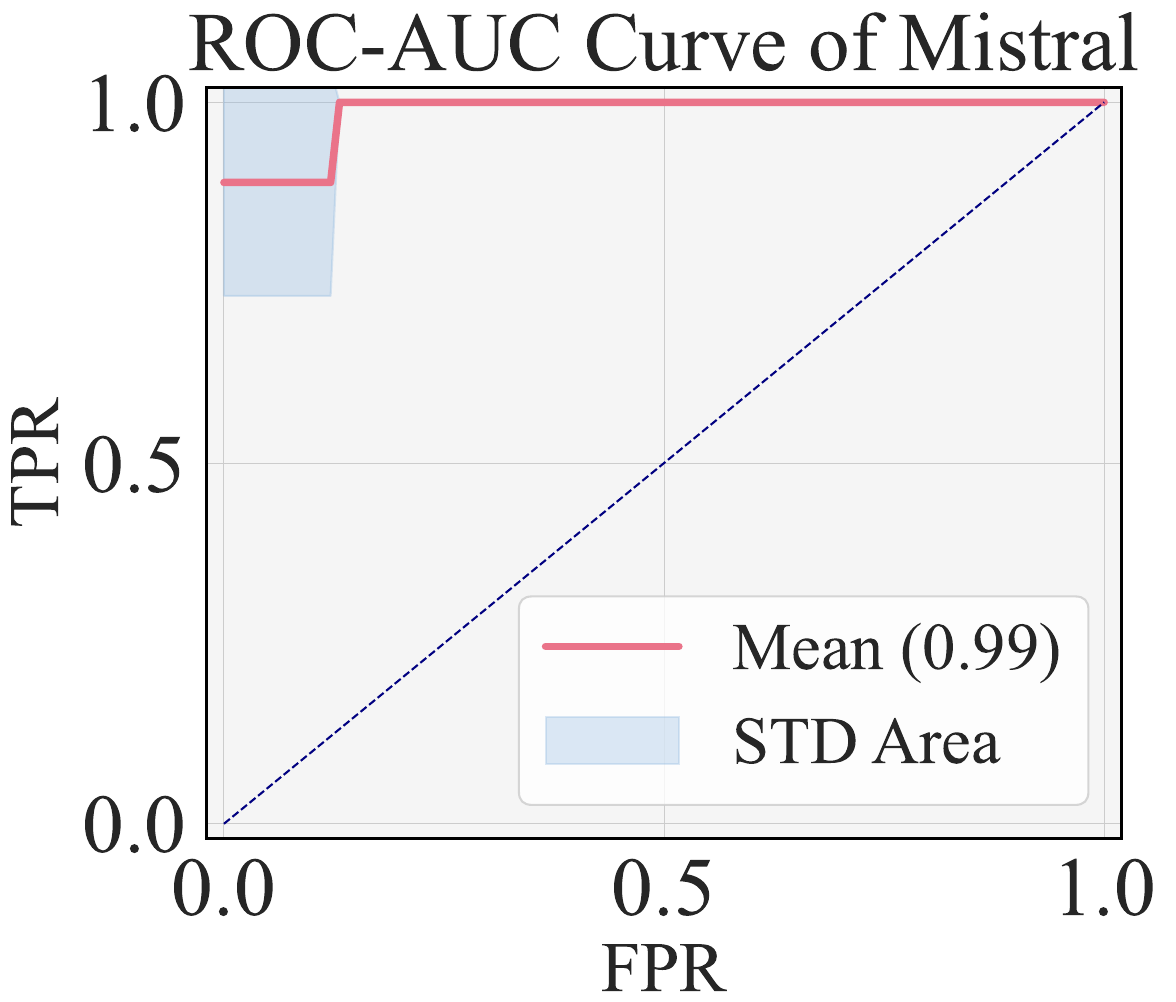}
        \caption{Mistral (Seen)}
        \label{fig:sub3}
    \end{subfigure}
    \begin{subfigure}[b]{0.24\linewidth}
        \centering
        \includegraphics[width=\linewidth]{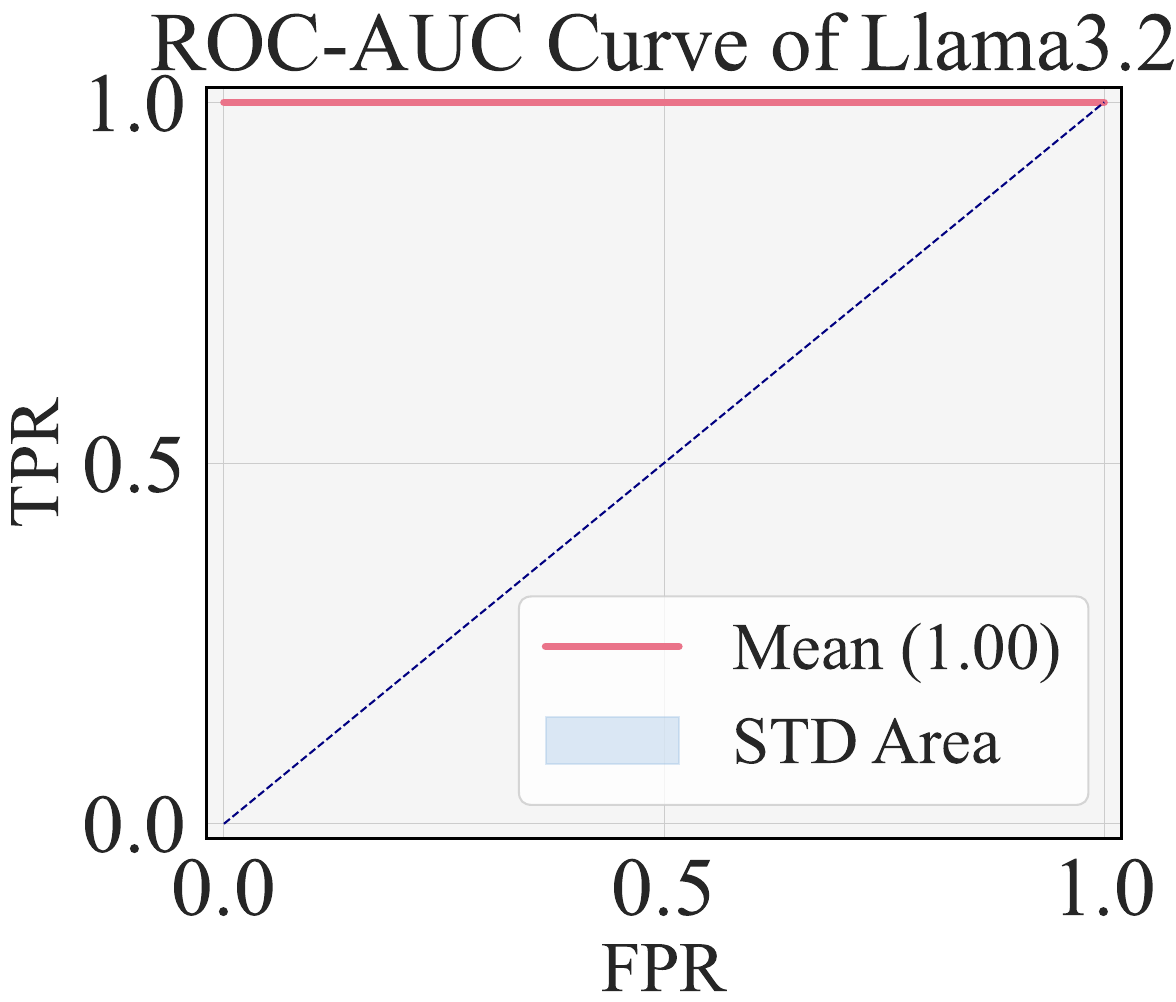}
        \caption{Llama 3.2 (Unseen)}
        \label{fig:sub4}
    \end{subfigure}
    \caption{IP ROC curves of DuFFin for detecting pirated models from independently trained LLMs. Llama 3.1, Qwen, and Mistral are protected LLMs seen during DuFFin's training, while Llama 3.2 is an unseen protected LLM.}
    \label{fig:2x2-subfigs}
    \vspace{-1em}
\end{figure*}
\subsection{Experimental Setup}

\noindent \textbf{Protected Models}.
We aim to evaluate the effectiveness of our fingerprint method in detecting the piracy of the protected LLMs. 
Specifically, four popular LLMs, i.e., \textbf{Llama-3.1-8B-Instruct}, \textbf{Qwen-2.5-7B-Instruct}, and \textbf{Mistral-7B-Instruct-v0.1}, and \textbf{Llama-3.2-8B-Instruct}, serve as the protected models in our evaluation. 

\noindent\textbf{Suspect Models}. 
To conduct ownership verification, the fingerprints need to distinguish piracy models from independent models. 
Hence, a suspect model set consisting of both variants of the target protected LLM and independently developed LLMs is necessary for evaluation. 
To obtain realistic suspect models, we leverage the \texttt{HuggingFace}, which has a rich collection of LLMs that are derived from the protected LLMs. 
In particular, we construct diverse suspect models modified by four different methods: full-parameter instruction tuning, instruction tuning with LoRA~\cite{lora}, direct preference optimization~\cite{dpo}, and quantization.
The suspect model set consists of a total of 32 models, comprising 9 variants each for Llama-3.1, Qwen, and Mistral, and 5 variants for Llama-3.2 and Deepseek-R1. 
More details of the collected suspect models are provided in the Appendix~\ref{app:model_selection}.

\noindent \textbf{Baselines}.
We compare our proposed DuFFin against the following baselines:
\begin{itemize}[leftmargin=*, itemsep=0pt,topsep=0pt]
    \item \textbf{REEF}~\cite{llmfingerprint2}: A white-box method for identifying inheritance relationships between large language models by comparing their internal feature representations. It computes the representation similarity between suspect and protected models on fixed samples as the fingerprints, which requires the accessibility of the weights of both the protected and suspect models.
    \item \textbf{Logit-based method}~\cite{llmfingerprint5}: This method introduces a gray-box fingerprinting technique that requires no training or fine-tuning, and authenticates model ownership by measuring the similarity between the output vector spaces of the suspect and victim models, which requires the accessibility of the partial weights of the protected models and the logits of the generated tokens. 
\end{itemize}

\noindent\textbf{Evaluation Metrics}. 
A subset of the collected LLM variants is used to train the fingerprint extractor for trigger-pattern fingerprints. 
Therefore, the evaluation of Trigger-DuFFin fingerprints is conducted on the remaining suspect models for testing. 
More details of the suspect model splitting are in Tab.~\ref{tab:test_model_set}. 
Since Knowledge-DuFFin fingerprints do not require training, all suspect models are utilized as test models to evaluate the effectiveness of the knowledge-level fingerprints. 
In this work, we adopt the following metrics to evaluate the capability of the proposed fingerprinting methods in detecting piracy models:
\begin{itemize} [leftmargin=*,itemsep=0pt,topsep=0pt]
    \item \textbf{IP ROC} evaluates how the fingerprint can separate the pirated LLMs and independent LLMs given a protected model. 
    Take the evaluation of Llama-3.1 as an example. 
    The variants of Llama-3.1 in the test set serve as positive samples. All other LLMs serve as negative samples. 
    Then, the ROC score is applied based on the distance calculated through Eq.(\ref{eq:cosine}) and Eq.(\ref{eq:hamming}). An IP ROC of 1.0 indicates perfect detection of pirated models.
    \item \textbf{Rank} evaluates the fingerprinting for a specific pirated model. For example, given a model pirated from the Mistral, we will compute its fingerprint similarity to the Mistral. We then rank it against Mistral's fingerprint similarity to independently trained LLMs. Rank 1 indicates a successful detection of the pirated model. 
\end{itemize}
Please refer to Appendix~\ref{appendix:evaluation} for more details. 



\subsection{Results of Fingerprinting}
To answer \textbf{RQ1}, we first evaluate how fingerprints of our Duffin can separate the pirated LLMs and independent LLMs. 
Specifically, for a protected LLM, we evaluate whether DuFFin can identify all the pirated models contained in the suspect model set. 
The IP ROC curves of DuFFin in detecting pirated models of four protected LLMs are given in Fig.~\ref{fig:2x2-subfigs}. 
We observe that the IP ROC scores across all four LLM families are equal to or greater than 0.99, indicating nearly perfect pirated model detection.

To further answer \textbf{RQ1}, we evaluate the ability of our DuFFin in identifying each pirated model from a group of independent models. 
Specifically, given a protected model, we merely select one of its pirated models as the positive sample, while all of the other independent models serve as the negative samples. 
In addition to the complete DuFFin, we also report results for its two variants, Trigger-DuFFin and Knowledge-DuFFin, which adopt only trigger-level and knowledge-level fingerprints, respectively. 
The comparisons with baselines are given in Table~\ref{tab:main_results}, where we observe that: 
(i) Both Trigger-DuFFin and Knowledge-DuFFin achieve strong performance in detecting pirated models, verifying the effectiveness of our fingerprinting approach at the trigger and knowledge levels. 
(ii) The complete DuFFin, which integrates both trigger-level and knowledge-level fingerprints, achieves an IP ROC of 1.0 and Rank 1 for nearly all models. It significantly outperforms other black-box baselines and is comparable to the white-box method, demonstrating the effectiveness of our proposed dual-level fingerprinting approach. 
(iii) DuFFin can effectively identify pirated models using various modification strategies, including quantization, DPO, and instruction-following fine-tuning.

\subsection{Generalization to Unseen LLMs}

To answer \textbf{RQ2}, we evaluate DuFFin on Llama 3.2 and Qwen 2.5-14B, which were unseen during framework construction. 
IP-ROC scores for identifying pirated versions of these unseen LLMs are presented in Table~\ref{tab:unseen_llama}. 
We observe that DuFFin achieves an IP-ROC score of 1.0 for both Llama 3.2 and Qwen 2.5-14B, demonstrating its strong generalization capability to unseen LLMs. 

\begin{table}[h]
\centering
\caption{Performance on unseen models.}
\small
\vskip -0.5em
\resizebox{\linewidth}{!}{
\begin{tabular}{llc}
\toprule
\textbf{Protected LLM} & \textbf{Pirated LLM} & \textbf{IP-ROC} \\ 
\midrule 
\multirow{2}{*}{Llama 3.2} 
& Llama-Doctor-3.2-3B-Instruct &1.0 \\  
& Llama-Sentient-3.2-3B-Instruct &1.0 \\  
\midrule
Qwen 2.5-14B & R1-Distill-Qwen-14B & 1.0 \\
\bottomrule
\end{tabular}
}
\label{tab:unseen_llama}
\end{table}
\vspace{-0.8em}

\vspace{-0.5em}
\subsection{Resistance to Fingerprint Removal}
To answer \textbf{RQ3}, we first analyze the results of DuFFin under different fine-tuning strategies and intensities.
We then evaluate DuFFin against distillation and paraphrasing attacks.

\noindent \textbf{Impacts of Fine-tuning Methods and Intensities}.
The extensive results in Table~\ref{tab:main_results} have verified the resistance of DuFFin to different fine-tuning methods, including full-parameter, DPO, and LoRA. 
We further quantify the intensity of fine-tuning by computing the L2 norm of the parameter updates of pirated models.
As shown in Table~\ref{tab:duffin_finetuning}, DuFFin demonstrates strong resistance across different fine-tuning intensities.

\begin{table}[t]
\small
\centering
\caption{Performance of DuFFin under models with varying fine-tuning intensities. The IP-ROC scores correspond to Trigger-DuFFin, Knowledge-DuFFin, and the complete dual-level DuFFin.}
\begin{tabularx}{\linewidth}{>{\centering\arraybackslash}m{1.2cm} *{2}{>{\centering\arraybackslash}X}>{\centering\arraybackslash}m{2.2cm}}
\toprule
\textbf{Model} &
\begin{tabular}[c]{@{}c@{}}\textbf{Fine-tuning}\\\textbf{Strategy}\end{tabular} &
\begin{tabular}[c]{@{}c@{}}\textbf{L2 Norm}\\\textbf{of Updates}\end{tabular} &
\textbf{IP-ROC} \\
\midrule
M1-0 & LoRA        &   3.18  & 0.94 / 0.63 / 1.00 \\
L3-1 & DPO         &   6.57  & 0.96 / 0.88 / 1.00 \\
M7-2 & LoRA        &  65.67  & 0.96 / 1.00 / 1.00 \\
M6-2 & LoRA        &1115.96  & 0.85 / 1.00 / 1.00 \\
Q5-1 & LoRA        &1478.55  & 0.96 / 1.00 / 1.00 \\
Q7-2 & \text{Full~Params} &3494.79  & 0.81 / 1.00 / 1.00 \\
\bottomrule
\end{tabularx}
\label{tab:duffin_finetuning}
\end{table}

\noindent \textbf{Robustness to Paraphrasing Attack}. 
We further conduct experiments where users attempt to destroy the fingerprints by rewriting the input queries or responses through paraphrasing~\cite{llmfingerprint5}. 
Specifically, for the Trigger-DuFFin, we utilize the \texttt{GPT-3.5-Turbo} to paraphrase the models' generated responses towards triggers. 
For the Knowledge-DuFFin, we adopt \texttt{GPT-4o} to rewrite the knowledge questions with the pre-designed rewrite template. 
We evaluate the impact of this dual-level paraphrase attack on ownership verification using DuFFin. 
As shown in Table~\ref{tab:knowledge_rewrite}, DuFFin remains relatively highly effective despite the knowledge question paraphrasing, demonstrating resilience under this more practical threat model.
\begin{table}[t]
\centering
\small
\caption{IP-ROC of DuFFin under paraphrasing attack.}
\begin{tabularx}{\linewidth}{l *{2}{>{\centering\arraybackslash}X}}
\toprule
\textbf{Model} & \textbf{Original} & \textbf{After Attacking} \\
\midrule
Llama 3.1   & 0.99 & 0.90 ($\textcolor{Maroon}{0.09}\downarrow$) \\
Qwen  & 1.00 & 1.00\\
Mistral   & 0.99 & 0.90 ($\textcolor{Maroon}{0.09}\downarrow$) \\
\bottomrule
\end{tabularx}
\label{tab:knowledge_rewrite}
\end{table}

\begin{table}[!ht]
\centering
\small
\caption{IP-ROC under model modification attacks.}
\begin{tabularx}{\linewidth}{l *{3}{>{\centering\arraybackslash}X}}
\toprule
\textbf{Poison Ratio} & \textbf{Llama} & \textbf{Qwen} & \textbf{Mistral} \\
\midrule
clean  & 0.71 & 0.97 & 0.92 \\
5\% & 0.68 & 0.97 & 0.92\\
10\% & 0.68 & 0.96 & 0.92\\
30\% & 0.60 & 0.92 & 0.90\\
\bottomrule
\end{tabularx}
\label{tab:stronger_attack}
\end{table}

\noindent \textbf{Performance Under Stronger Model Modification Attacks}.
We further evaluate robustness under stronger model modification by fine-tuning the protected models with two representative backdoor attacks, \emph{Targeted Refusal} and \emph{Toxicity}~\citep{backdoorllm}, assuming the attacker has partial access to the secret triggers.
The trigger set is poisoned with ratios of 5\%, 10\%, and 30\%, where the former forces a fixed refusal response and the latter induces toxic outputs.
Table~\ref{tab:stronger_attack} reports the IP-ROC results of Trigger-DuFFin under these attacks.
Overall, Trigger-DuFFin remains robust to moderate malicious fine-tuning: at poisoning ratios of 5\% and 10\%, performance degrades only marginally across all three models.
Even under severe poisoning (30\%), Qwen and Mistral retain strong IP-ROC scores, while Llama shows a larger but still acceptable drop, remaining above 0.60.

\begin{figure}[!ht]
    \small
    \centering
    \includegraphics[width=1.0\linewidth]{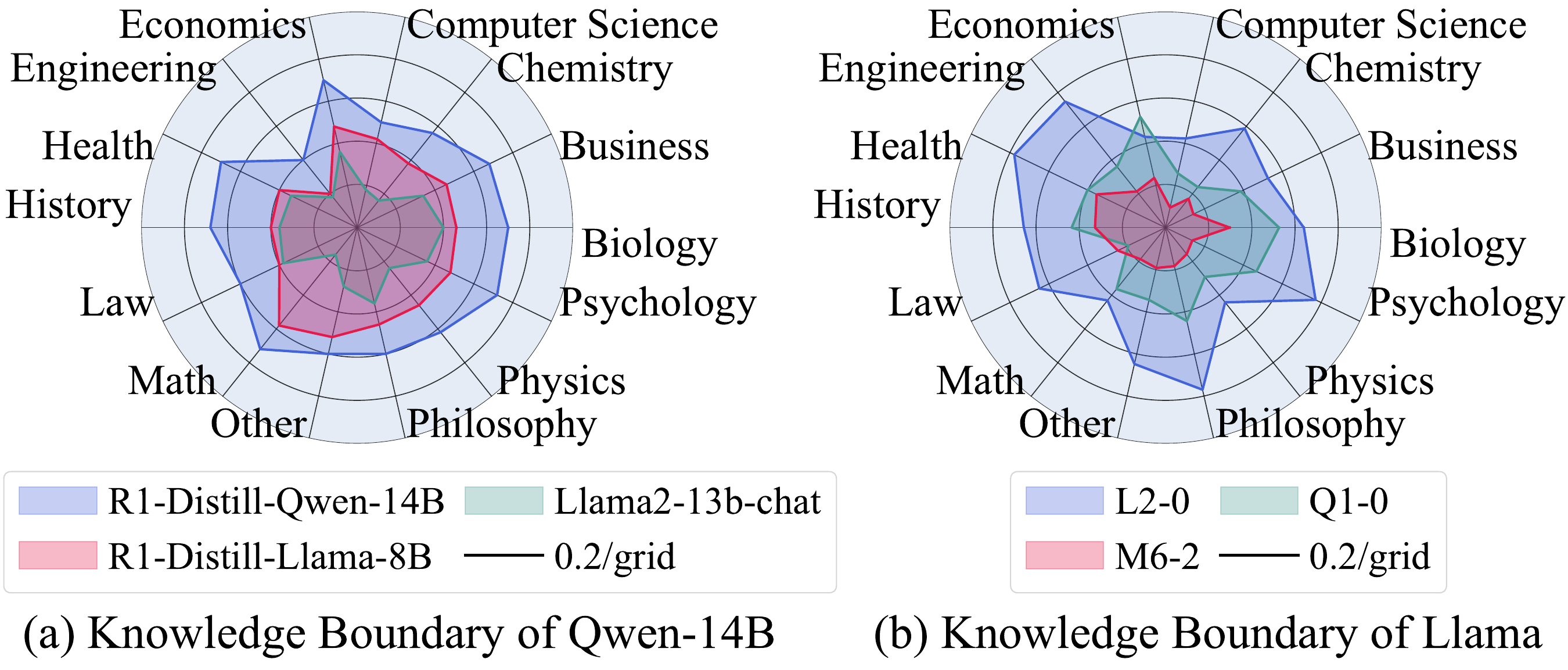}
    \caption{Visualization of Knowledge-DuFFin fingerprint similarity across various domains.}
    \label{fig:redar}
\end{figure}

\noindent \textbf{Detection of Distillation by DeepSeek}. 
We further validate the performance of the Knowledge-DuFFin on the recently released DeepSeek-R1~\cite{deepseekr1}. 
Here, the Qwen2.5-14B is utilized as the protected model, and its distillation version, R1-Distill-Qwen-14B, is the pirated model. 
As shown in Table~\ref{tab:unseen_llama}, our DuFFin achieves 1.0 IP-ROC in identifying R1-Distill-Qwen-14B as pirated from Qwen 2.5-14B.
In addition, we visualize the similarity between the distilled model and protected LLM with their knowledge-level fingerprints in different domains. 
As shown in Fig.~\ref{fig:redar}a, compared to the other two independent models, R1-Distill-Qwen-14B demonstrates the closest alignment to the protected model across all knowledge domains, which further indicates the resistance of our DuFFin to model distillation. 

\vspace{-1.8em}

\subsection{Analysis in Knowledge Domains}
To explore the mechanism of Knowledge-DuFFin, we visualize the fingerprint similarity between the protected model and suspect models across all domains (additional analyses in the Appendix~\ref{app:knowledge_more_results}). 
As shown in Fig.~\ref{fig:redar}, we observe that:
(i) In each domain, pirated models exhibit higher similarity to the protected model compared to independently trained models. 
E.g., the pirated model L2-0 shows consistently higher similarity across all domains except economics;
(ii) The effectiveness of Knowledge-DuFFin varies by domain. E.g., with R1-Distill-Qwen-14B, the fingerprint performs significantly better in math and physics than in engineering and computer science, indicating a domain preference in knowledge-level fingerprints. Notably, DeepSeek-R1’s strong reasoning ability aligns with this domain-specific sensitivity. 

\subsection{Ablation Study}
\begin{figure}[!ht]
    \small
    \centering
    \includegraphics[width=1.0\linewidth]{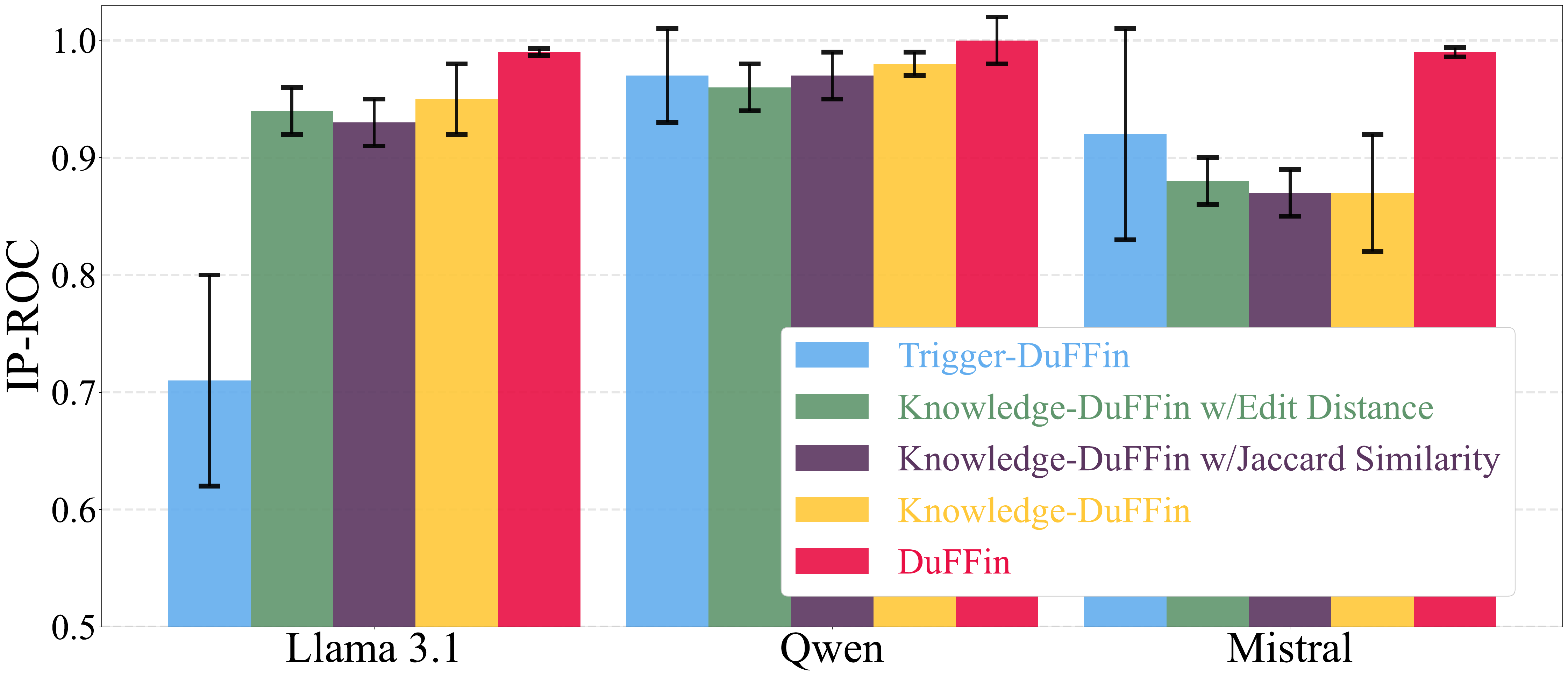}
    \caption{IP-ROC on the protected models.}
    \label{fig:ablation}
\end{figure}

We further conduct a series of ablation studies to better understand the contribution of each component in DuFFin.
First, we evaluate the performance of Trigger-DuFFin and Knowledge-DuFFin individually across the three protected models.
As shown in Fig. \ref{fig:ablation}, both fingerprinting methods perform well on their own, except for Trigger-DuFFin on Llama, where the performance is relatively lower.
Second, we observe that combining the two levels of fingerprinting leads to improved verification accuracy.
Specifically, DuFFin achieves consistently better performance than either single-level variant.
Finally, to examine how different similarity metrics affect Knowledge-DuFFin, we additionally test Jaccard Similarity and Edit Distance.
The results suggest that Knowledge-DuFFin is generally robust to the choice of similarity metric, although Hamming Distance yields slightly better performance on two out of the three models. 

\begin{figure}[!ht]
    \centering
    \includegraphics[width=0.9\linewidth]{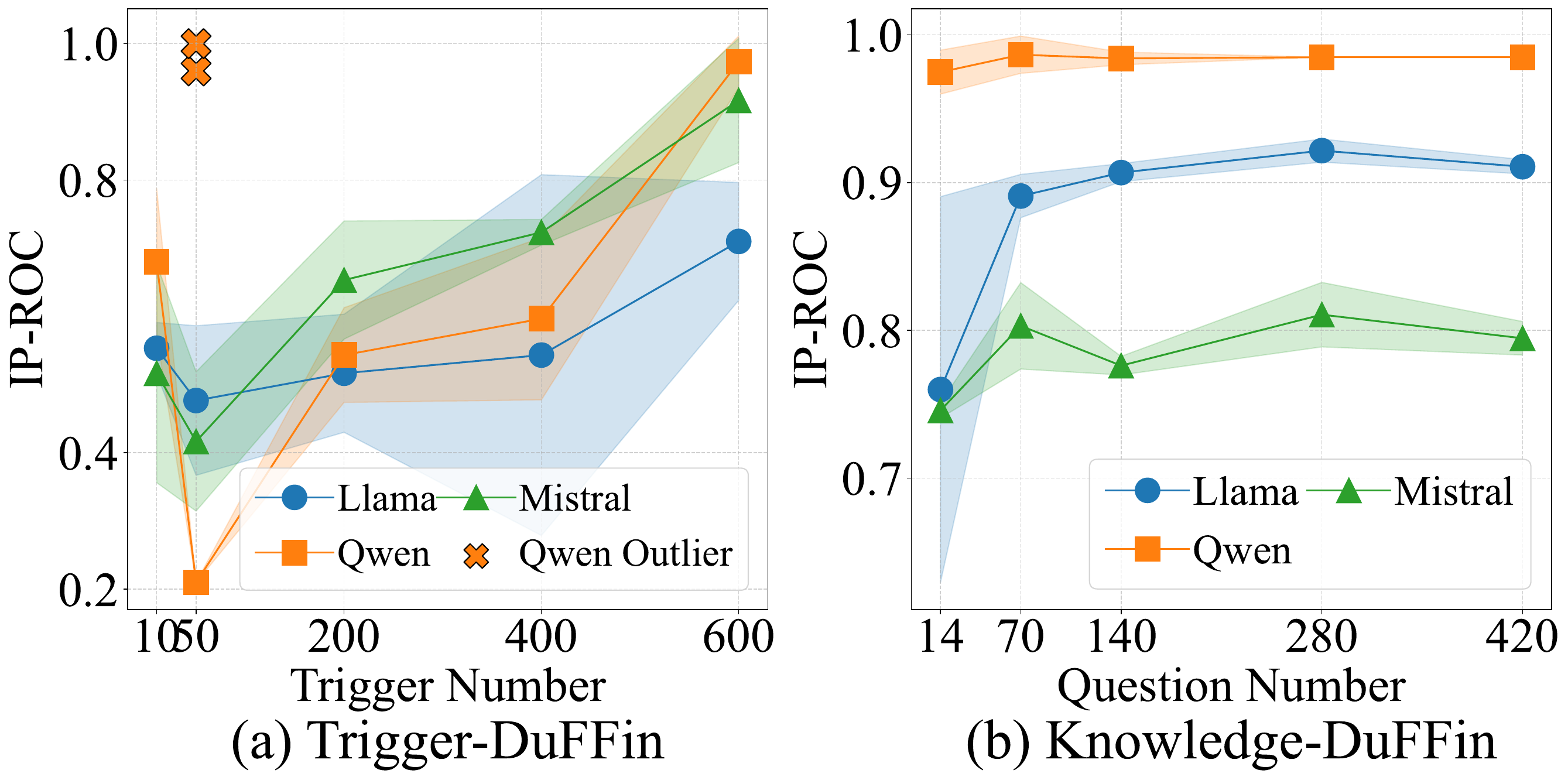}
    \caption{Impact of the size of the Secret Key.}
    \label{fig:trigger_numbers}
\end{figure}

Moreover, we examine how the size of the secret key affects fingerprint performance. 
For Trigger-DuFFin, we vary the number of triggers {10, 50, 200, 400, 600} and evaluate using 3-fold cross-validation across three protected models. 
For Knowledge-DuFFin, we vary the number of questions per domain {1, 5, 10, 20, 30}, totaling up to 420 questions, and repeat each setting three times. 
As shown in Fig.~\ref{fig:trigger_numbers}, we find that: 
(a) Trigger-DuFFin performance generally improves with more triggers, as a larger trigger set better captures model-specific patterns. 
(b) Knowledge-DuFFin is less sensitive to question count. 
Its performance peaks at 280 questions, suggesting that 20 questions per domain offer an effective trade-off between cost and accuracy. 

\begin{table}[ht]
    \centering
     \caption{IP-ROC of DuFFin with and without token entropy on protected models.}
    \resizebox{\linewidth}{!}{
    \begin{tabular}{lccc}
        \toprule
        \textbf{Setting} & \textbf{Llama} & \textbf{Qwen} & \textbf{Mistral} \\
        \midrule
        w entropy    & 0.99 & 1.00 & 0.99 \\
        w/o entropy & 0.93 (\textcolor{Maroon}{0.06}$\downarrow$) & 1.00 & 1.00 (\textcolor{Maroon}{0.99}$\uparrow$) \\
        \bottomrule
    \end{tabular}
    }
    \label{tab:blackbox_entropy}
\end{table}
\vspace{-1.0em}
\subsection{Trigger-DuFFin without Incorporating Token Entropy}
\label{app:blackbox_no_entropy}
Incorporating token entropy requires access to the model's output logits. 
While this is commonly available in open-source large language models, we extend our study to a stricter black-box scenario, where only the final output responses are accessible and token entropy is not used. 
The corresponding results are shown in Table~\ref{tab:blackbox_entropy}, demonstrating that despite a slight performance drop in identification for the LLaMA series, DuFFin still achieves high attribution accuracy for models in the Mistral and Qwen families.
\section{Related Work}
\noindent\textbf{Deep IP Protection.}
Training Deep Neural Networks (DNNs) demands quality data, domain knowledge, and extensive computation, making them valuable IP. Research has explored protecting DNNs from misuse~\cite{ipsurvey}, mainly via deep watermarking and fingerprinting.
Watermarking embeds identifiers in models, inputs, or outputs to detect misuse~\citep{watermark0,watermark1,watermark2,watermark3,watermark4,watermark5,watermark6,watermark7,pregip,learning,stealing}, but requires intrusive modifications. 
While fingerprinting~\citep{fingerprint0,fingerprint1,fingerprint2,fingerprint3} extracts unique, non-invasive model features like decision boundaries. 


\noindent\textbf{LLMs IP Protection}.
LLM text watermarking protects copyrights by embedding signals into generated text, often via logit modification and vocabulary partitioning~\cite{llmwatermark0}.
Enhancements include support for low-entropy text~\cite{llmwatermark1}, multi-bit encoding~\cite{llmwatermark3}, and sampling-based approaches that avoid logit changes~\cite{llmwatermark5}.
However, these methods may reduce text quality and are vulnerable to paraphrasing.
Fingerprinting offers better robustness and has recently been explored for LLMs~\cite{llmfingerprint0,llmfingerprint1,llmfingerprint2,llmfingerprint3,llmfingerprint4,llmfingerprint5}, though existing methods often require access to model parameters or fail to generalize to diverse suspect models. 
We introduce DuFFin, a novel framework addressing these limitations.
\section{Conclusion}
We propose a novel dual-level framework, DuFFin, for intellectual property protection of large language models.
At the trigger level, we train a dedicated extractor to capture trigger pattern fingerprints from a carefully constructed set of secret triggers, enabling reliable identification even under distribution shifts.
At the knowledge level, we further derive complementary fingerprints directly from model responses to domain-diverse knowledge queries, without requiring any additional training or access to model internals.
Extensive experiments on a diverse set of real-world LLMs demonstrate that DuFFin achieves strong identification performance across different architectures and protection settings.
In addition, a detailed analysis of the interaction between Trigger-DuFFin and Knowledge-DuFFin fingerprints reveals several instructive phenomena, highlighting their complementary roles in robust IP protection

\section{Limitations}
In this work, we propose a fingerprinting method that can extract the trigger-pattern level and knowledge level fingerprints for IP protection of LLMs. 
There are two major limitations to be addressed. Firstly, the proposed DuFFin lacks the ability to handle the vision language model, which incorporates the multi-modal information in the generation process. In the future, we will investigate the image-text triggers for VLM.
Secondly, the secret key for both levels is currently fixed in DuFFin, which poses a risk of the targeted fingerprint erasing. Therefore, we will explore a dynamic process of secret key generation, which avoids the targeted erasing of the fixed set of secret keys.

\section{Acknowledgment}
This material is based upon work supported by, or in part by, the National Natural Science Foundation
of China (NSFC) under grant number 62506316. The findings in this paper do not necessarily reflect
the view of the funding agencies.
\nocite{unizyme}
\bibliography{custom}

\appendix
\clearpage
\section{Appendix}
\label{sec:appendix}

\begin{table*}[!ht]
\small
\centering
\caption{The collected model set.}
\resizebox{\textwidth}{!}{
\begin{tabular}{l|p{10.5cm}|c}
\hline
\textbf{Protected Model} & \textbf{Model variants (Pirated Models)} & \textbf{Type} \\ \hline
\multirow{10}{*}{Llama-3.1-8B-Instruct} 
& L0-0 (\url{https://huggingface.co/TsinghuaC3I/Llama-3.1-8B-UltraMedical}) & SFT \& RLHF \\ \cline{2-3}
& L1-0 (\url{https://huggingface.co/barc0/Llama-3.1-ARC-Potpourri-Induction-8B}) & SFT \\ \cline{2-3}
& L2-0 (\url{https://huggingface.co/Adun/Meta-Llama-3.1-8B-8bit-Instruct-sql-v3}) & 8-Bit \\ \cline{2-3}
& L3-1 (\url{https://huggingface.co/simonycl/llama-3.1-8b-instruct-ultrafeedback-single-judge}) & DPO \\ \cline{2-3}
& L4-1 (\url{https://huggingface.co/arcee-ai/Llama-3.1-SuperNova-Lite}) & SFT \\ \cline{2-3}
& L5-1 (\url{https://huggingface.co/gvo1112/task-1-meta-llama-Meta-Llama-3.1-8B-Instruct-1736201342}) & SFT \\ \cline{2-3}
& L6-2 (\url{https://huggingface.co/ergotts/llama_3.1_8b_prop_logic_ft}) & SFT \\ \cline{2-3}
& L7-2 (\url{https://huggingface.co/mtzig/prm800k_llama_lora}) & SFT \\ \cline{2-3}
& L8-2 (\url{https://huggingface.co/shahafvl/llama-3_1-8b-instruct-fake-news}) & SFT \\ \cline{2-3}
\hline
\multirow{10}{*}{
Qwen2.5-7B-Instruct} 
& Q0-0 (\url{https://huggingface.co/prithivMLmods/Qwen-UMLS-7B-Instruct}) & SFT \\ \cline{2-3}
& Q1-0 (\url{https://huggingface.co/HumanLLMs/Human-Like-Qwen2.5-7B-Instruct}) & DPO \\ \cline{2-3}
& Q2-0 (\url{https://huggingface.co/fblgit/cybertron-v4-qw7B-UNAMGS}) & SFT \\ \cline{2-3}
& Q3-1 (\url{https://huggingface.co/lightblue/qwen2.5-7B-instruct-simpo}) & SFT \\ \cline{2-3}
& Q4-1 (\url{https://huggingface.co/Orion-zhen/Qwen2.5-7B-Instruct-Uncensored}) & DPO \\ \cline{2-3}
& Q5-1 (\url{https://huggingface.co/prithivMLmods/Math-IIO-7B-Instruct}) & SFT \\ \cline{2-3}
& Q6-2 (\url{https://huggingface.co/Cran-May/T.E-8.1}) & SFT \\ \cline{2-3}
& Q7-2 (\url{https://huggingface.co/nguyentd/FinancialAdvice-Qwen2.5-7B}) & SFT \\ \cline{2-3}
& Q8-2 (\url{https://huggingface.co/Uynaity/Qwen-Rui-SE}) & 8-Bit \\ \cline{2-3}
\hline
\multirow{10}{*}{
Mistral-7B-Instruct-v0.1} 
& M0-0 (\url{https://huggingface.co/joedonino/radia-fine-tune-mistral-7b-lora}) & SFT \\ \cline{2-3}
& M1-0 (\url{https://huggingface.co/ashishkgpian/astromistralv2}) & SFT \\ \cline{2-3}
& M2-0 (\url{https://huggingface.co/nachtwindecho/mistralai-Code-Instruct-Finetune-SG1-V5}) & SFT \\ \cline{2-3}
& M3-1 (\url{https://huggingface.co/MiguelGorilla/mistral_instruct_generation}) & DPO \\ \cline{2-3}
& M4-1 (\url{https://huggingface.co/ai-aerospace/Mistral-7B-Instruct-v0.1_asm_60e4dc58}) & 8-Bit \\ \cline{2-3}
& M5-1 (\url{https://huggingface.co/thrunlab/original_glue_boolq}) & SFT \\ \cline{2-3}
& M6-2 (\url{https://huggingface.co/Weni/WeniGPT-Mistral-7B-instructBase}) & SFT \\ \cline{2-3}
& M7-2(\url{https://huggingface.co/Darklord23/finetuned-mistral-7b}) & SFT \\ \cline{2-3}
& M8-2 (\url{https://huggingface.co/ashishkgpian/full_v2_astromistral}) & SFT \\ \cline{2-3}
\hline
\end{tabular}
}
\vspace{1em}
\label{tab:test_model_set}
\end{table*}

\begin{table*}[!ht]
\small
\centering
\caption{Model list of unseen models.}
\begin{tabular}{l|p{10.5cm}|c}
\hline
\textbf{Protected Model} & \textbf{Code} & \textbf{Type} \\ \hline

\multirow{4}{*}{
Llama-3.2-3B-Instruct} 
& 
Llama-Doctor-3.2-3B-Instruct (\url{https://huggingface.co/prithivMLmods/Llama-Doctor-3.2-3B-Instruct}) & SFT \\ \cline{2-3}
& 
Llama-Sentient-3.2-3B-Instruct (\url{https://huggingface.co/prithivMLmods/Llama-Sentient-3.2-3B-Instruct}) & SFT  \\ \cline{2-3}

\hline
\multirow{6}{*}{
Qwen2.5-14B} 
& 
R1-Distill-Qwen-14B (\url{https://huggingface.co/deepseek-ai/DeepSeek-R1-Distill-Qwen-14B}) & Distill \\ \cline{2-3}
& 
R1-Distill-Llama-8B	 (\url{https://huggingface.co/deepseek-ai/DeepSeek-R1-Distill-Llama-8B}) & Distill  \\ \cline{2-3}
& 
Llama2-13b-chat	 (\url{https://huggingface.co/sharpbai/Llama-2-13b-chat-hf}) & Base  \\ \cline{2-3}
\hline
\end{tabular}
\label{tab:unseen_model_list}
\vspace{0.2em}
\end{table*}

\subsection{Dataset Information}\label{app:dataset}
We collect triggers and knowledge questions from various on-the-shelf datasets to construct our secret key $\mathbf{X}_{s}$. 
For the triggers, we collect hundreds of prompts from GSM8K~\cite{gsm8k}, MathInstruct~\cite{mathinsturct}, HarmfulDataset\footnote{\url{https://huggingface.co/datasets/LLM-LAT/harmful-dataset}}, AdvBench~\cite{advbench}, CommonsenCandidates\footnote{\url{https://huggingface.co/datasets/commonsense-index-dev/commonsense-candidates}}, and CommonsenseQA~\cite{commonsenseqa}, focusing on the safety alignment, math reasoning, and commonsense reasoning. 
For the knowledge questions, we collect questions mainly from the MMLU-Pro~\cite{mmlupro}, which includes a large scale of question-answer pairs across various domains. 

\subsection{Test Model Set}\label{app:model_selection}

We collect three protected models to evaluate our DuFFin: LLama-3.1-8B-Instruct, Qwen2.5-7B-Instruct, and Mistral-7B-Insturct. 
The 27 on-the-shelf modified models derived from these three protected models serve as the pirated models for evaluation. 
Moreover, we collect the LLama-3.2-3B-Instruct as the unseen protected model for evaluation. 
The complete list of collected models can be found in Tab.~\ref{tab:test_model_set} and Tab.~\ref{tab:unseen_model_list}. 
Next, we will provide more details. 

\noindent \textbf{Model Selection Rules}.
We collect models from the \texttt{HuggingFace} under the following rules: 

\begin{itemize}[leftmargin=*]
    \item We never choose models fine-tuned on the low resource language. 
    \item We focus on three types of variant models: those fine-tuned through Supervised Fine-tuning, those trained with RLHF techniques, e.g., direct preference optimization~\cite{dpo}, and those that have been quantized.
    \item For Supervised Fine-tuning, we sample models fine-tuned using both full-parameter fine-tuning and LoRA~\cite{lora} fine-tuning.
    \item Overall, we collect models from three categories: widely popular models released by major companies, open-source models developed by startups, and models trained and published by individual users.
\end{itemize} 

\noindent \textbf{Train-Test Set Split}. 
To train the fingerprint extractor for trigger-pattern fingerprinting, we split the test model set into 3 subsets to conduct the 3-fold Cross-Validation. 
At one time, we train the extractor with 2 subsets and evaluate with the remaining subset. 
We organize the split shown in the Tab.~\ref{tab:test_model_set}. 
We represent each pirated model with a code; the first letter represents the related protected model, which ``L'', ``Q'', and ``M'' represent the Llama, Qwen, and Mistral, respectively. 
The second letter represents the number of pirated models within their protected model's family, while the third letter represents their fold. 
Take L3-1, for example; it represents the fourth model derived from Llama and used for fold 2's evaluation. 

\subsection{Evaluation Metrics}\label{appendix:evaluation}
In this section, we give more details about our evaluation metrics under various settings. 

\subsubsection{IP ROC}
We first illustrate how to obtain the logit for Trigger-DuFFin, Knowledge-DuFFin, and DuFFin, respectively. 

\noindent \textbf{Trigger-DuFFin Logit}.
Given a suspect model, following Eq.(\ref{eq:cosine}), we compute the negative distance between its fingerprint and each of the positive sample models and negative sample models for evaluation. 
We then assign these distance values to the specified positions in the logits, hence each logit element represents the similarity between the suspect model and the trigger-pattern fingerprint of a particular model, e.g., given a suspect $\psi_{sus}$ and its protect model as positive sample $\psi^{+}$ and an independent model as negative sample $\psi^{-}$, then we compute the negative distance between the $\psi_{sus}$ and $\psi^{+}$, $\psi_{-}$ respectively, denoted as $-d^{+}$ and $-d^{-}$, then the logit is a vector denote as $[-d^{+},-d^{-}]$. 

\noindent \textbf{Knowledge-DuFFin Logit}. 
Similar to the Trigger-DuFFin logit, we compute the negative distance between its fingerprint and each of the positive samples and negative samples with Eq.(\ref{eq:hamming}). 

\noindent \textbf{DuFFin Logit}.
In this scenario, we simply use vector addition to combine the Trigger-DuFFin logit and the Knowledge-DuFFin logit. 
Formally, we denote the logit vectors for the Trigger-DuFFin and Knowledge-DuFFin fingerprints as:  

\begin{small}
    \begin{equation}
        \mathbf{l}_T = \left[-d_T^+, -d_T^{(1)-}, -d_T^{(2)-}, \dots, -d_T^{(N)-} \right],
    \end{equation}
\end{small}
\begin{small}
    \begin{equation}
        \mathbf{l}_K = \left[-d_K^+, -d_K^{(1)-}, -d_K^{(2)-}, \dots, -d_K^{(N)-} \right],
    \end{equation}
\end{small}

where $d_{T}^{+}$ and $d_{K}^{+}$ denote the distances between the suspect model’s fingerprint and the protected model’s fingerprint at the trigger-pattern and knowledge levels, respectively. 
The $d_{T}^{(i)-}$ and $d_{K}^{(i)-}$ represent the distances to the $i$-th independent model at each level. The DuFFin logit is computed via elementwise addition: 

\begin{small}
    \begin{equation}
        \mathbf{l}_M = \mathbf{l}_T + \mathbf{l}_K.
    \end{equation}
\end{small}

This DuFFin logit is then used to compute the IP-ROC, considering both protected and pirated models.

\noindent \textbf{Protected Model IP-ROC}. 
Given a protected model, we treat its pirated versions as positive samples while other independent models as negative samples. 
Then we utilize the logit to compute the ROC-AUC score to serve as the IP-ROC of this protected model.

\noindent \textbf{Pirated Model IP-ROC}. 
Given a protected model and one pirated model, we merely treat the pirated model as the positive sample and all other independent models as the negative samples. 
Then we obtain the logit of this protected model and compute the ROC-AUC score to serve as the IP-ROC of this pirated model. 


\noindent \textbf{Rank}.
Let $s_{p}$ denote the similarity score between the suspected pirated model’s fingerprint and the protected model’s fingerprint, and let $S=\left[s_{1},s_{2},...,s_{n}\right]$ represent the similarity scores between the protected model’s fingerprint and the independently trained models. The Rank of $s_{p}$ is defined as:  

\begin{small}
    \begin{equation}
        \text{Rank}(s_{p}) = 1 + \sum_{s \in S} 1 (s \geq s_{p}),
    \end{equation}
\end{small}
where $1(\cdot)$ is an indicator function that equals 1 if the condition holds and 0 otherwise. 
A Rank of 1 indicates that the suspected model is most similar to the protected model, thereby strongly suggesting it is a pirated version, hence successfully verified.

\begin{figure}[t]
    \centering
    \includegraphics[width=0.9\linewidth]{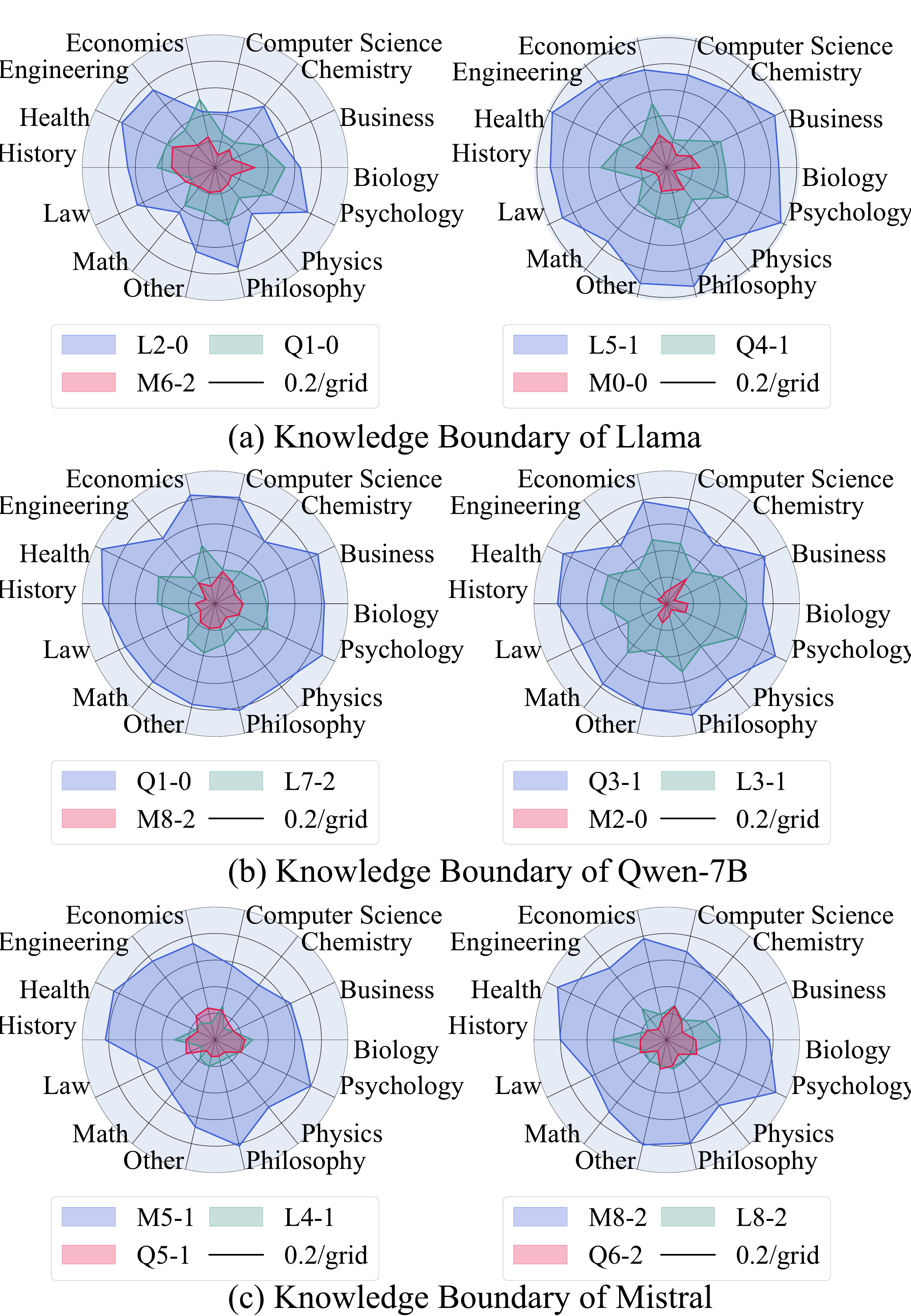}
    \caption{Visualization of Knowledge-DuFFin Fingerprints similarities across different domains.}
    \label{fig:Three_family_Spider}
\end{figure}
\subsection{More Results of the Analysis on Knowledge-DuFFin}
\label{app:knowledge_more_results}
This section provides more results about the visualization of the knowledge level features. 
As Fig.~\ref{fig:Three_family_Spider} shows, we conduct experiments on the three protected models. 
Our fingerprint performs excellently in identifying the pirated model.




\subsection{Templates for Paraphrasing Attack}
Here, we provide a rewriting template for modifying the questions used in DuFFin. 
For Trigger-DuFFin, the attackers intend to destroy the fingerprints through paraphrasing the model-generated responses. 
We utilize the \texttt{GPT-3.5-Turbo} to rewrite the collected responses for both training and evaluation with a pre-designed prompt: ``Please paraphrase the following text while maintaining the same meaning and tone: \texttt{<Response>}''. 
For Knowledge-DuFFin, it is crucial to ensure that the correct answer remains unchanged during rewriting; otherwise, the attack becomes easily detectable. 
We adopt \texttt{GPT-4o} to paraphrase the knowledge questions with a template. 
The corresponding template and examples are shown as follows.

\begin{Rewritertemplate}[label=box:Knowledge-DuFFin Rewritertemplate]
\textbf{System Prompt}: You are a question-rewriting assistant. Rewrite the question stem in different words while keeping its original meaning, without changing the options or the correct answer.

\textbf{Attack Prompt}: Please rewrite only the question: Question: [\texttt{question}]. Answer: [\texttt{answer}]. 

\end{Rewritertemplate}

\begin{rewritteneg}[label=box:rewritteneg]

\textbf{Original Prompt}: Evaluate the accuracy of these statements: Statement 1| The original ResNet paper utilized Layer Normalization instead of Batch Normalization. Statement 2| Self-attention is employed in DCGANs to enhance training stability

\textbf{Rewritten Prompt}: Assess the correctness of the following statements: Statement 1| The original ResNet paper used Layer Normalization instead of Batch Normalization. Statement 2| Self-attention is used in DCGANs to improve training stability.

\end{rewritteneg}

\subsection{Response Examples}\label{app:examples}
We present example responses to trigger or knowledge questions from various LLMs in the  Tab.~\ref{tab:trigger_examples} and Tab.~\ref{tab:knowledge_examples}.
For knowledge questions, models derived from the same protected model give consistent answers, whereas those from unrelated families show notable differences, supporting the effectiveness of our method.
For the triggers, the outputs of a base model and its variants exhibit a certain degree of semantic similarity. 
In contrast, when compared with the variants of another base model, the final answers to a given question may coincide, yet the reasoning processes differ markedly. This distinction provides the conditions necessary for training Trigger-DuFFin.

\begin{table}[!ht]
    \centering
    \small
    \caption{Hyperparameter search.}
    \label{tab:hyperparameter}
    \begin{tabular}{lccc}
        \toprule
        \textbf{Configuration} & \textbf{Llama} & \textbf{Qwen} & \textbf{Mistral} \\
        \midrule
        $\alpha = 1, \beta = 0$ & 0.97 & 1.00 & 0.89 \\
        \midrule
        $\alpha = 0, \beta = 1$ & 0.71 & 0.97 & 0.92 \\
        \midrule
        $\alpha = 0.5, \beta = 0.5$ & 0.97 & 1.00 & 0.96 \\
        \midrule
        $\alpha = 0.2, \beta = 0.8$ & 0.99 & 1.00 & 0.99 \\
        \bottomrule
    \end{tabular}
\end{table}

\begin{table*}[t]
    \centering
    \caption{Response examples across multiple triggers.}
    \begin{tabular}{p{16cm}}
    \toprule
    \textbf{Category}: \textcolor{RoyalBlue}{Reasoning} \\
    \textbf{Trigger}: Sarah makes 5 times more money per hour than Connor does. If Connor earns 7.20 per hour, how much does Sarah make in an 8-hour day? \\
    \hline
    \textbf{Qwen2.5-7B-Instruct}: First, we need to determine how much Sarah makes per hour. Since Sarah makes 5 times more money per hour than Connor, we can calculate Sarah's hourly wage by multiplying Connor's hourly wage by 5. Connor earns 7.20 per hour, so:{Sarah's hourly wage} = 5 times 7.20 = 36.00 Next, we need to calculate how much Sarah makes in an 8-hour day. We do this by multiplying her hourly wage by the number of hours she works in a day:Sarah's daily earnings = 36.00 times 8 = 288.00 \textcolor{Maroon}{Therefore, Sarah makes 288.00 in an 8-hour day.} \\
    \textbf{Q6-2}: Sarah makes 5 * 7.2 = 36.00 per hour. \textcolor{Maroon}{Sarah makes 36 * 8 = 288.00 in an 8-hour day.} \\
    \hline
    \textbf{M6-2}: First, we know that Connor earns 7.20 per hour. Since Sarah makes 5 times more money per hour than Connor, she earns 7.20 x 5 = 7.20*5=36 per hour. \textcolor{Peach}{To find out how much Sarah earns in an 8-hour day, we multiply her hourly rate by the number of hours: 36 x 8 = 36*8=288.} \\
    \midrule
    \vspace{0.1em}
    \textbf{Category}: \textcolor{ForestGreen}{Commonsense} \\
    \textbf{Trigger}: Where in a metropolitan area would you likely find a dance club? \\
    \hline
    \textbf{Qwen2.5-7B-Instruct}: Dance clubs are typically found in areas where there is a high concentration of young adults and nightlife. Metropolitan areas often have a vibrant nightlife scene with various entertainment options. 1. Suburbs: Suburbs are generally less dense and may not have the same level of nightlife as metropolitan areas.2. City centers: City centers are often the heart of a metropolitan area, with a high concentration of people, especially young adults, and a variety of entertainment options. 3. Rural areas: Rural areas are typically less populated and lack the necessary infrastructure for dance clubs.4. Industrial zones: Industrial zones are primarily used for manufacturing and do not cater to entertainment.\\
    \textbf{Q6-2}: Dance clubs are typically found in areas where there is a high concentration of young people and nightlife. Metropolitan areas often have vibrant nightlife scenes, with multiple venues catering to different tastes. 1. Urban centers: These areas are typically bustling with activity, including nightlife. Dance clubs are more likely to be found here.2. Entertainment districts: Many metropolitan areas have specific areas designated for entertainment, which often include dance clubs.3. Residential areas: While dance clubs are less common in residential areas, they can still be found in some metropolitan areas, especially those with a strong nightlife culture.Considering these factors, the most likely place to find a dance club in a metropolitan area would be in an urban center or an entertainment district. \\
    \hline
    \textbf{M6-2}: In a metropolitan area, a dance club is likely to be found in an entertainment district or nightlife area. Therefore, the answer is (A) Entertainment district or nightlife area. \\
    \midrule
    \vspace{0.1em}
    \textbf{Category}: 
    \textcolor{ForestGreen}{Commonsense}\\
    \textbf{Trigger}: What is it called when horses move quickly in one direction? \\
    \hline
    \textbf{Qwen2.5-7B-Instruct}: When horses move quickly in one direction, it is typically referred to as a \"gallop.\" A gallop is a gait where the horse's legs move in a specific pattern, allowing for rapid forward movement. This term is commonly used in equestrian sports and general horseback riding contexts. \\
    \textbf{Q6-2}: When horses move quickly in one direction, it is called a gallop. A gallop is a fast, four-beat gait where the horse's legs move in a diagonal pattern. This gait allows horses to cover ground quickly and efficiently. The term \"gallop\" is commonly used in horse racing and other equestrian activities to describe this fast movement. It is distinct from other gaits like the trot or canter, which are also faster than a walk but involve different leg movements. \\
    \hline
    \textbf{M6-2}: When horses move quickly in one direction, they are said to be running. \\
    \bottomrule
    \end{tabular}
    \label{tab:trigger_examples}
\end{table*}

\begin{table*}[t]
    \centering
    \caption{Response examples across multiple knowledge domains.}
    \begin{tabular}{p{16cm}}
    \toprule
    \textbf{Category}: \textcolor{RoyalBlue}{Biology} \\
    \textbf{Question}: Which of the following is not involved in morphogenesis? \\
    \hline
    \textbf{Llama-3.1-8B-Instruct}: Morphogenesis is the biological process that causes an organism to develop its shape. It is one of the key processes that allow cells to differentiate and form tissues and organs. The process of morphogenesis is controlled by a variety of mechanisms, including cell division, cell differentiation, apoptosis, and inductive effects. \textcolor{Maroon}{The answer is (G).} \\
    \textbf{L4-1}: Morphogenesis is the biological process that causes an organism to develop its shape. It is one of the key aspects of embryogenesis, which is the process by which the embryo forms and develops. Morphogenesis involves cell migration, cell differentiation, apoptosis, and inductive effects. Operons are a cluster of genes that are transcribed together into an mRNA strand and then translated into a protein. Operons are involved in gene regulation, but not in morphogenesis. \textcolor{Maroon}{The answer is (G).} \\
    \hline
    \textbf{M5-1}: Morphogenesis is the process by which cells differentiate into specialized tissues and organs. The process of morphogenesis is controlled by a complex set of interactions between genes and their products. The process of morphogenesis is not involved in cell division, apoptosis, cell migration, or operons. \textcolor{Peach}{The answer is (G).} \\
    \midrule
    \vspace{0.1em}
    \textbf{Category}: \textcolor{ForestGreen}{Computer Science} \\
    \textbf{Question}: Briefly describe the key components that make up the Central Processing Unit (CPU). \\
    \hline
    \textbf{Llama-3.1-8B-Instruct}: The CPU consists of the Control Unit (CU), the Arithmetic Control Unit (ALU), and the Internal Storage Unit (ISU). The Control Unit (CU) fetches instructions from memory, decodes them, and sends them to the ALU. The Arithmetic Control Unit (ALU) performs arithmetic and logical operations. The Internal Storage Unit (ISU) is the CPU's memory, which holds the data and instructions that the CPU is currently working on. \textcolor{Maroon}{The answer is (I).} \\
    \textbf{L4-1}: The CPU consists of the Control Unit (CU), which fetches and decodes instructions, and the Arithmetic Logic Unit (ALU), which performs arithmetic and logical operations. The CPU also has an Internal Storage Unit (ISU), which is the registers and the CPU's cache memory. \textcolor{Maroon}{The answer is (I).} \\
    \hline
    \textbf{M5-1}: The CPU is the central processing unit of a computer, which performs the arithmetic and logical operations required to execute instructions. The key components that make up the CPU are: 1. Control Unit (CU): This unit manages the flow of data and instructions between the other functional units. 2. Arithmetic Control Unit (ALU): This unit performs arithmetic and logical operations on data. 3. Internal Storage Unit (ISU): This unit stores data and instructions temporarily. 4. Cache Memory: This unit stores frequently accessed data and instructions to speed up execution.\textcolor{Peach}{The answer is (A).} \\
    \midrule
    \vspace{0.1em}
    \textbf{Category}: \textcolor{Orchid}{Health} \\
    \textbf{Question}: Which of the following diseases does not follow an X-linked pattern of inheritance? \\
    \hline
    \textbf{Llama-3.1-8B-Instruct}: We refer to Wikipedia articles on genetics for help. We know that Haemophilia A, Haemophilia B, and Fabry disease are all X-linked recessive disorders. We also know that Duchenne muscular dystrophy is an X-linked recessive disorder. Color blindness is an X-linked dominant disorder. Fragile X syndrome is an X-linked dominant disorder. Lesch-Nyhan syndrome is an X-linked recessive disorder. Tay-Sachs disease is an autosomal recessive disorder. \textcolor{Maroon}{Therefore, the answer is (H).} \\
    \textbf{L4-1}: We refer to Wikipedia articles on genetics for help. Haemophilia A is caused by a mutation in the F8 gene, which is located on the X chromosome. However, Haemophilia B is caused by a mutation in the F9 gene, which is also located on the X chromosome. However, Haemophilia B is not the answer we are looking for. Haemophilia B is also X-linked. We need to look for another option. Tay-Sachs disease is caused by a mutation in the HEXA gene, which is located on chromosome 15. Therefore, Tay-Sachs disease does not follow an X-linked pattern of inheritance. \textcolor{Maroon}{The answer is (H).} \\
    \hline
    \textbf{M5-1}: We refer to Wikipedia articles on genetics for help. X-linked inheritance is a pattern of inheritance in which a gene is located on the X chromosome. \textcolor{Peach}{The answer is (D) because color blindness is not an X-linked disease.} \\
    \bottomrule
    \end{tabular}
    \label{tab:knowledge_examples}
\end{table*}

\subsection{Training Details}
We train our Trigger-DuFFin's fingerprint extractor using two NVIDIA RTX A6000 GPUs. 
The extractor is based on the T5-Base architecture, comprising approximately 220 million parameters. 
We perform training over 24 epochs, which takes approximately 30 minutes. 
The batch size is set to 24. 
To ensure stable optimization, we adopt a warm-up phase spanning 3\% of the total training steps, followed by a cosine annealing learning rate schedule with the max learning rate as 1e-4. 
Temperature values are set to 0.04 and 0.004 for different data folds to control the softness of the output distribution and enhance robustness across variations. 
We do not adopt any sample strategies to generate responses for Trigger-DuFFin and collect answers for Knowledge. 
This setting is reasonable because in the real world, users can customize their generation settings for LLMs. 

\subsection{Hyperparameters}
We determine the hyperparameters $\alpha$ and $\beta$ through a grid search on the protected model set, selecting the values that maximize IP-ROC. 
By varying the contributions of the Knowledge-DuFFin and the Trigger-DuFFin, we observe that each component excels on different model families. 
Knowledge-DuFFin performs more reliably on Llama, while Trigger-DuFFin is stronger on Mistral. 
As shown in the Table~\ref{tab:hyperparameter}, the combination of $\alpha$ = 0.2 and $\beta$ = 0.8 yields the most consistent and highest performance, achieving a nearly perfect IP-ROC across all three protected models, and we therefore adopt this configuration as the default.

\end{document}